\documentclass[journal]{IEEEtran}
\usepackage{mathrsfs}
\usepackage{amsfonts}
\usepackage{ifpdf}
\usepackage{cite}
\ifCLASSINFOpdf
\usepackage[pdftex]{graphicx}
\else \fi
\usepackage[cmex10]{amsmath}
\usepackage{array}
\usepackage{mdwmath}
\usepackage{mdwtab}
\usepackage{eqparbox}
\usepackage[tight,footnotesize]{subfigure}
\usepackage{algorithmic}
\usepackage{algorithm}
\usepackage{amsmath}
\usepackage{epsfig}
\usepackage{ae}
\usepackage{amssymb}
\usepackage{times}
\usepackage{amsmath}   
\usepackage{booktabs}  
\usepackage{threeparttable} 
\usepackage{multirow}       
\usepackage{xcolor}
\usepackage{graphicx}
\usepackage{subfigure}
\usepackage{makecell}
\newtheorem{theorem}{Theorem}
\newtheorem{lemma}{Lemma}

\begin{document}

\title{Long-term Scheduling and Power Control for Wirelessly Powered Cell-free IoT}

\author{Xinhua Wang, Xiaodong Wang, \textit{Fellow, IEEE}, Alexei Ashikhmin, \textit{Fellow, IEEE}

\thanks{Copyright (c) 2015 IEEE. Personal use of this material is permitted. However, permission to use this material for any other purposes must be obtained from the IEEE by sending a request to pubs-permissions@ieee.org.

Xinhua Wang is with the College of Electrical Engineering, Qingdao University, Qingdao, 266071 China (e-mail: xhwang@qdu.edu.cn).

Xiaodong Wang is with the Electrical Engineering Department, Columbia University, New York, NY 10027 USA (e-mail: wangx@ee.columbia.edu).
          
Alexei Ashikhmin is with the Nokia Bell Labs, Murray Hill, NJ 07974 USA (e-mail: alexei.ashikhmin@nokia-bell-labs.com).

}
}

\maketitle

\begin{abstract}
We investigate the long-term scheduling and power control scheme for a wirelessly powered cell-free Internet of Things (IoT) network which consists of distributed access points (APs) and large number of sensors. In each time slot, a subset of sensors are scheduled for uplink data transmission or downlink power transfer. Through asymptotic analysis, we obtain closed-form expressions for the harvested energy and the achievable rates that are independent of random pilots. Then, using these expressions, we formulate a long-term scheduling and power control problem to maximize the minimum time average achievable rate among all sensors, while maintaining the battery state of each sensor higher than a predefined minimum level. Using Lyapunov optimization, the transmission mode, the active sensor set, and the power control coefficients for each time slot are jointly determined. Finally, simulation results validate the accuracy of our derived closed-form expressions and reveal that the minimum time average achievable rate is boosted significantly by the proposed scheme compare with the simple greedy transmission scheme.
\end{abstract}

\begin{IEEEkeywords}
Cell-free IoT, Lyapunov optimization, long-term scheduling and power control, max-min fairness, wireless power transfer.
\end{IEEEkeywords}
\section{Introduction}

\IEEEPARstart{N}{owdays}, Internet of Things (IoT) has become a ubiquitous technology with wide applications on transportation, healthcare, agriculture, and other aspects of our daily life \cite{A. Al-Fuqaha, F. Javed}. The limited energy storage of IoT devices \cite{Z. Chu} and massive connectivity \cite{Y. Xiao} are two main challenges to hinder the proliferation of IoT technologies, and have attracted intensive research interests in recent years \cite{K. Mikhaylov}.

Wireless power transfer (WPT) is regarded as a promising technology to tackle the energy shortage problem of IoT devices. WPT has been mainly considered for three scenarios: energy broadcasting without information transmission \cite{L. Liu, S. Kashyap}, simultaneous wireless information and power transfer (SWIPT) \cite{T. D. Ponnimbaduge Perera, J. Huang_a}, and the wireless powered communication network (WPCN) which is more suitable for IoT \cite{S. Bi, J. Chen}. In WPCN, IoT devices first harvest RF energy from the downlink, and then transmit data to APs through uplink. The main challenge of WPCN is the heavy path loss in both downlink and uplink. Although multiple antennas can improve the efficiency of WPT \cite{T. A. Khan}, the performance of cell-boundary users is still poor. 

Recently, cell-free massive MIMO with distributed and cooperative APs has been proposed to improve both the spectral and energy efficiencies. Compared with cellular massive MIMO, the heavy path-loss of the cell boundary terminals can be avoided since the distances between the terminals and served APs are smaller. With the help of cell-free massive MIMO with a user-centric architecture \cite{H. Q. Ngo_a, E. Nayebi}, Wang \textit{et al.} proposed a wirelessly powered cell-free IoT scheme with jointly optimized downlink and uplink power control coefficients \cite{Xinhua Wang}. Compared with WPT in small cells and collocated massive MIMOs, the WPT efficiency of the cell-free IoT has been improved significantly.

In cell-free massive MIMO systems, random pilots are wildly adopted since it is impossible to allocate finite orthogonal pilots to massive amount of sensors \cite{Xinhua Wang, S. Rao}. Intuitively, the non-orthogonality of random pilots will reduce the accuracy of channel estimation. Then, the degradation of channel estimation can further decrease the efficiency of WPT and data transmission. Therefore, user scheduling is critical to satisfy the massive connections in wirelessly powered cell-free IoT. The traditional scheduling methods usually focus on improving the instantaneous network performance \cite{M. Mollanoori,X. Wang_ta}, which may cause some sensors with bad channel conditions never scheduled. Therefore, it is more appropriate to design a scheduling strategy to improve the long-term system performance. Lyapunov optimization method is an effective way to improve the infinite horizon average objective function without predicting the future state \cite{M. J. Neely, S. Pan}, which has been used to design scheduling strategies in some works \cite{S. Pan, D. Zhai}. In particular, with the help of Lyapunov optimization, Zhai \textit{et al.} designed an energy-efficient user scheduling for NOMA-based IoT networks to minimize the time average power consumption while satisfying the time average rate requirements for all devices\cite{ D. Zhai}. Through optimizing energy beamforming, Choi \textit{et al.} minimized the time average power consumption of the AP while meeting the given time average rate requirements for all nodes in a WPCN \cite{K. Choi}. It is challenging to jointly design the long-term scheduling and power control scheme for a wirelessly powered network, since we need to strike a balance between the WPT and data transmission and to guarantee the fairness between users.

\emph{\textbf{Contributions}}: Our contributions in this work are three-fold:
	\begin{itemize}
		\item We provide an asymptotic analysis to obtain closed-form expressions for the harvested energy and the achievable rates, that are independent of random pilot sequences.
		\item To serve massive amount of sensors, a long-term scheduling and power control optimization problem is formulated to maximize the minimum time average achievable rate, while maintaining the battery state of each sensor higher than a predefined level. By solving the problem, the transmission mode (energy harvesting or data transmission), the active sensor set, the power control coefficients for each time slot are jointly determined. 
        \item  Under the Lyapunov optimization framework, the long-term problem is transformed into a sequence of optimization problems of minimizing the Lyapunov drift plus penalty for each time slot, which can be solved efficiently using our proposed low-complexity methods. Simulation results reveal that our scheduling and power control scheme can boost the minimum time average achievable rate significantly.	
	\end{itemize}
	The remainder of this paper is organized as follows. In Section II we describe system model and formulate the problem. The asymptotic analysis is provided in Section III. In Section IV, we transform the long-term scheduling and power control problem into Lyapunov optimization. The algorithm for solving the uplink and downlink sub-problems in each time slot are described in Section V. Simulation results are given in Section VI. Finally Section VII concludes the paper.

\section{System Model and Outline of Results}
\subsection{System Model} 
We consider a wirelessly powered cell-free IoT network \cite{Xinhua Wang} with \(L\) APs and a set of randomly distributed single-antenna sensors denoted as $\mathcal{K}$ with $|\mathcal{K}|=K$. Each AP equipped with \(N\) antennas is connected to a central processing unit (CPU) via a perfect back-haul network. In each time slot $t$, only a subset $\mathcal{K}_a^{(t)}$ with $|\mathcal{K}_a^{(t)}|=K_a$ are scheduled as active sensors, while the remaining sensors are inactive. As shown in Fig. 1, each time slot of duration $\Delta$ seconds contains \(T_c\) OFDM symbols, in which \(\tau\) OFDM symbols are used for channel estimation, while the remaining \(T_c-\tau\) symbols are used for downlink WPT if \(\delta^{(t)}=1\) or uplink data transmission if \(\delta^{(t)}=0\), where
\begin{align}\label{trans_mode}
    \delta^{(t)}=\{0,1\},~t=0,1,2,\cdots
\end{align}
indicates the transmission mode with $\bar{\delta}^{(t)}=1-{\delta}^{(t)}$.

Let \({\pmb \theta}^{(t)}=\left[\theta_1^{(t)},\cdots,\theta_K^{(t)}\right]^T\) 
denote the sensor states in time slot $t$ with 
\begin{align}\label{schedule_set}
    |{\pmb \theta}^{(t)}|_1=K_a, ~\mbox{and}~ \theta_k^{(t)}\in \{0,1\}, k\in \mathcal{K},
\end{align}
i.e., the set of active sensors is \(\mathcal{K}_a^{(t)}=\{k:\theta_k^{(t)}=1\}\). The channel between the \(k\)-th sensor and the \(n\)-th antenna of the \(l\)-th AP in time slot $t$ is
 \[g^{(t)}_{(l,n),k}=\sqrt{\beta_{l,k}}h^{(t)}_{(l,n),k} \]
 where \(\beta_{l,k}\) is the large-scale fading coefficient, which depends on the location and is assumed known to the APs. \(h_{(l,n),k}^{(t)}\sim \mathcal{CN}(0,1)\) represents the small-scale fading coefficient, which remains invariant in each time slot, but varies independently from one slot to another. The channel between the \(l\)-th AP and the \(k\)-th sensor is denoted as
\[{\pmb{g}}_{l,k}^{(t)}=\left[{g}_{(l,1),k}^{(t)},\cdots,{g}_{(l,N),k}^{(t)}\right]^T\in \mathbb{C}^{N \times 1}, \] 
while the channel between the \(n\)-th antenna of \(l\)-th AP and all sensors is denoted as
	\[\pmb{g}_{(l,n)}^{(t)}=\left[g_{(l,n),1}^{(t)}, \cdots,g_{(l,n),K}^{(t)}\right]^T\in \mathbb{C}^{K \times 1}. \]
 \begin{figure}
	\centering \scalebox{1}{\includegraphics[width=\columnwidth]{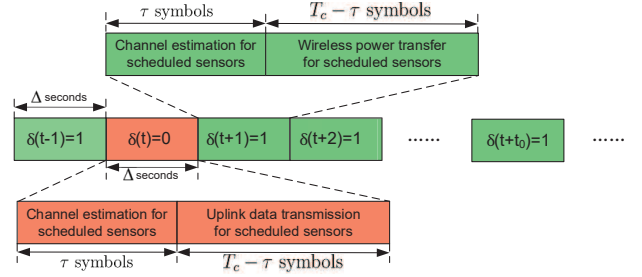}}
	\centering \caption{Scheduling for cell-free IoT.}
\end{figure}
\subsection{Channel Estimation Phase}
At the channel estimation phase in each time slot, all \(K_a\) active sensors simultaneously transmit their pilot sequences to APs. Let \(\pmb{\Psi}=\left[\pmb{\psi}_{1},\cdots,\pmb{\psi}_{K}\right]\in \mathbb{C}^{\tau\times K}\), where \(\pmb{\psi}_k \sim \mathcal{CN}\left( {\pmb 0}, \frac{1}{\tau}{\pmb I}_{\tau}\right)\) denotes the pilot sequence of the \(k\)-th sensor. At the $t$-th slot, the received pilots at the \(n\)-th antenna of the \(l\)-th AP is given by
\begin{align}\label{received_pilots}
\pmb{y}_{(l,n)}^{(t)}&=\sqrt{\tau\rho_p}\sum_{k=1}^K\theta_{k}^{(t)}{g}_{(l,n),k}^{(t)}\pmb{\psi}_k+\pmb{w}_{(l,n)}^{(t)}\nonumber \\
	&=\sqrt{\tau\rho_p}\pmb{\Psi}{\pmb \Theta}^{(t)}\pmb{g}_{(l,n)}^{(t)}+\pmb{w}_{(l,n)}^{(t)}\in \mathbb{C}^{\tau\times 1},
	\end{align}
where \(\pmb{w}_{(l,n)}^{(t)}\sim \mathcal{CN}({\pmb 0},\sigma^2 \pmb{I}_{\tau})\) is the additive noise, \(\rho_p\) is the pilot transmit power, and ${\pmb \Theta}^{(t)}={\rm diag}({\pmb \theta}^{(t)})$. Using the received pilots $\pmb{y}_{(l,n)}^{(t)}$, the LMMSE channel estimate \(\hat{\pmb{g}}_{(l,n)}^{(t)}\) is given by
\begin{align*}
\hat{\pmb{g}}_{(l,n)}^{(t)}&=[{\pmb A}_l^{(t)}]^H{\pmb y}_{(l,n)}^{(t)},
\end{align*}
which implies that the $k$-th entry of \(\hat{\pmb{g}}_{(l,n)}\) is
\begin{align}\label{g_lnk_est}
{\hat g}^{(t)}_{(l,n),k}=[\pmb{a}^{(t)}_{l,k}]^H{\pmb y}_{(l,n)}^{(t)},
\end{align}
where
\begin{align}
\pmb{A}^{(t)}_{l}=\sqrt{E_p}\left(\pmb{I}_{\tau}+E_p\pmb{\Psi}\pmb{D}_l^{(t)}\pmb{\Psi}^H\right) ^{-1}\pmb{\Psi}\pmb{D}_l^{(t)},\nonumber
\end{align}
$E_p=\tau\rho_p/{\sigma^2}$, \(\pmb{D}_l^{(t)}=\text{diag}\left(\beta_{l,1},\cdots,\beta_{l,K}\right){\pmb \Theta}^{(t)}\), and $\pmb{a}_{l,k}^{(t)}$ is the $k$-th column of ${\pmb A}_l^{(t)}$. Define
\begin{align}\label{Z_l}
\pmb{Z}_l^{(t)}&=\pmb{I}_{\tau}+E_p\pmb{\Psi}\pmb{D}_l^{(t)}\pmb{\Psi}^H\nonumber\\
&=\pmb{I}_{\tau}+\sum\nolimits_{j\in \mathcal{K}_a^{(t)}}E_p\beta_{l,j}{\pmb \psi}_j{\pmb \psi}_j^H.
\end{align}
The mean squared value of the channel estimate ${\hat g}_{(l,n),k}^{(t)}$ is given by
\begin{align}\label{Gamma_lk}
	 \gamma_{l,k}^{(t)}&=\mathbb{E}\left[\left|{\hat g}_{(l,n),k}^{(t)}\right|^2\right]=\left[\mathbb{E}\left(\hat{\pmb{g}}_{(l,n)}^{(t)}[\hat{\pmb{g}}_{(l,n)}^{(t)}]^H\right)\right]_{kk}\nonumber \\
  &=\sqrt{\tau\rho_p}\beta_{l,k}\pmb{\psi}_k^H\pmb{a}^{(t)}_{l,k}\nonumber\\
  &=\tau\rho_p\beta^2_{l,k}\pmb{\psi}_k^H[\pmb{Z}_l^{(t)}]^{-1}\pmb{\psi}_k,~~ k\in \mathcal{K}_a^{(t)}.
	\end{align}
	
\subsection{Tranmission Phase}
 \subsubsection{Downlink WPT}
If $\delta^{(t)}=1$, during the remaining $T_c-\tau$ symbols, the APs jointly perform energy beamforming to the active sensors based on channel estimates, i.e., the transmitted signal from the \(l\)-th AP within the $\ell$-th OFDM symbol is 
\begin{align}\label{DL_signal}
    \pmb{x}_{l}^{(t,\ell)}=\sqrt{\rho_d}\sum_{j=1}^K \sqrt{\eta_{l,j}^{(t)}\theta_j^{(t)}}\left[{\hat{\pmb{g}}^{(t)}_{l,j}}\right]^* q_j^{(t,\ell)}, \ell=\tau+1,\cdots,T_c,
\end{align}
where \(q_j^{(t,\ell)} \sim \mathcal{CN}(0,1/(T_c-\tau))\) is the $\ell$-th symbol to the \(j\)-th sensor, and 
\begin{align}\label{down_cons}
    \eta_{l,j}^{(t)}\geq 0, \forall l,j,
\end{align} 
denote the downlink power control coefficients. The transmit power of each AP is constrained by
\begin{align}\label{AP_constraints1}
	    P_l^{(t)}=\sum_{\ell=\tau+1}^{T_c}\mathbb{E}\left[\left\|\pmb{x}_l^{(t,\ell)}\right\|^2\right]\leq N\rho_d, \forall l
	\end{align}
where \(N\rho_d\) is the maximum transmit power of each AP. Substituting (\ref{Gamma_lk}) and (\ref{DL_signal}) into (\ref{AP_constraints1}), we have 
\begin{align}\label{AP_constraints}
\sum_{k=1}^K \delta^{(t)}\theta_k^{(t)}{\eta}_{l,k}^{(t)}{\gamma}_{l,k}^{(t)} \leq 1,~\forall l.
\end{align}

Then, the received signal at the \(k\)-th sensor is
	\begin{align}\label{received_sensor}
	z_k^{(t,\ell)}&=\sum_{l=1}^L\left[\pmb{g}^{(t)}_{l,k}\right]^T\pmb{x}_{l}^{(t,\ell)}+v_k^{(t,\ell)},\ell=\tau+1,\cdots,T_c,
	\end{align}
	where \(v_k^{(t)}\sim \mathcal{CN}(0,~\sigma^2/(T_c-\tau))\) is the noise. The amount of harvested energy at the each active sensor during each time slot can be expressed as
		 \begin{align}\label{received_power}
\mathcal{E}_{k}^{(t)}&=(1-\alpha)\Delta \zeta\delta^{(t)}\theta_k^{(t)} (T_c-\tau)\mathbb{E}\left[\left|z_k^{(t,\ell)}\right|^2\right]
  \end{align}
where \(\alpha={\tau}/{T_c}\) and \(\zeta \in (0,~1)\) represents the energy conversion efficiency.

  \subsubsection{Uplink Data Transmission}
   If $\delta^{(t)}=0$, during the remaining $T_c-\tau$ symbols, \(K_a\) active sensors simultaneously deliver their data to APs. The received signal at the \(l\)-th AP within the $\ell$-th OFDM symbol is given by
   \begin{align*}
    \pmb{r}_{l}^{(t,\ell)}=\sqrt{\rho_u}\sum_{j=1}^K\sqrt{\xi_j^{(t)}\theta_j^{(t)}}\pmb{g}_{l,j}^{(t)}s_j^{(t,\ell)}+\pmb{n}_{l}^{(t,\ell)},
    \end{align*}
  where $s_j^{(t,\ell)}$ denotes the $\ell$-th uplink symbol of the \(j\)-th sensor with $\mathbb{E}\left[|s_j^{(t,\ell)}|^2\right]=1/{(T_c-\tau)}$, \(\rho_u\) is the maximum transmit power of each sensor, \(\pmb{n}^{(t,\ell)}_l\sim \mathcal{CN}({\pmb 0},~\frac{\sigma^2}{T_c-\tau}{\pmb I}_{N})\) is the additive noise, and \(\xi_k^{(t)}\) represents the power control coefficient of the $k$-th sensor with 
  \begin{align}\label{up_cons_1}
      0 \leq \xi_k^{(t)}\leq 1, k\in \mathcal{K}.
  \end{align}
  The energy consumption of each active sensor for data transmission in each time slot is
    \begin{align}\label{up_cons_2}
      E_k^{(t)}=(1-\alpha) \Delta \rho_u\theta_k^{(t)} \xi_k^{(t)}\leq b_k^{(t)}, k\in \mathcal{K},
  \end{align}
  where $b_k^{(t)}$ is the battery state of the $k$-th sensor at the beginning of time slot $t$. Each AP individually performs beamforing and then sends \(\hat{\pmb{g}}^H_{l,k}\pmb{r}_{l}^{(t,\ell)}\) to the CPU, which detects \(s_k^{(t,\ell)}\) using matched filtering (MF) as follows
	\begin{align*}
	&\hat{s}_k^{(t,\ell)}=\sum\limits_{l=1}^L[\hat{\pmb{g}}_{l,k}^{(t)}]^H\pmb{r}^{(t,\ell)}_{l}=\underbrace{\sqrt{\rho_u\xi_k^{(t)}}\sum\limits_{l=1}^L\mathbb{E}\left[[\hat{\pmb{g}}_{l,k}^{(t)}]^H\pmb{g}^{(t)}_{l,k}\right]}_{\mathcal{C}_1}s^{(t,\ell)}_k\nonumber \\&+\underbrace{\sqrt{\rho_u\xi_k^{(t)}}\sum\limits_{l=1}^L\left([\hat{\pmb{g}}_{l,k}^{(t)}]^H\pmb{g}_{l,k}^{(t)}-\mathbb{E}\left[[\hat{\pmb{g}}_{l,k}^{(t)}]^{H}\pmb{g}_{l,k}^{(t)}\right]\right)}_{\mathcal{C}_2}s_k^{(t,\ell)}\nonumber \\&+\underbrace{\sum_{j\neq k,j\in \mathcal{K}}\sqrt{\rho_u\xi_j^{(t)}\theta_j^{(t)}}\sum\nolimits_{l=1}^L[\hat{\pmb{g}}_{l,k}^{(t)}]^{H}\pmb{g}_{l,j}^{(t)}}_{\mathcal{C}_3}s_j^{(t,\ell)}+\underbrace{\sum\nolimits_{l=1}^L[\hat{\pmb{g}}_{l,k}^{(t)}]^{H}\pmb{n}^{(t,\ell)}_{l}}_{\mathcal{C}_4},
	\end{align*}
 where \(\mathcal{C}_1\) is the desired signal, and \(\mathcal{C}_2+\mathcal{C}_3+\mathcal{C}_4\) is the effective noise. Since \(\mathcal{C}_1\), \(\mathcal{C}_2\), \(\mathcal{C}_3\), and \(\mathcal{C}_4\) are uncorrelated, the achievable rate of the \(k\)-th sensor is lower bounded by \cite{H. Q. Ngo_a,E. Nayebi}
 \begin{align}\label{achievable_rate}
    R_k^{(t)}=(1-\alpha)\bar{\delta}^{(t)}\theta_k^{(t)}\log_2(1+\Gamma_k^{(t)}) ~~\mbox{bits/s/Hz},
\end{align}
where the effective SINR $\Gamma_k$ is
 \begin{align}\label{SINR}
	\Gamma_k^{(t)}=\frac{|\mathcal{C}_1|^2 }{\mathbb{E}\left[|\mathcal{C}_2|^2\right]+\mathbb{E}\left[|\mathcal{C}_3|^2\right]+\mathbb{E}\left[|\mathcal{C}_4|^2\right]},
	\end{align}
	where the expectation is taken over small-scale fading.
 \subsubsection{Battery Model} 
 Let $b_{\max}$ denote the capacity of the battery. Since $b_k^{(t)}-E_{k}^{(t)}\geq 0$ from (\ref{up_cons_2}), the battery state of the $k$-th sensor $b_k^{(t)}$ is updated at the beginning of each time slot according to 
 \begin{align}\label{bat_state}
 b_k^{(t+1)}=\min\{b_k^{(t)}-\bar{\delta}^{(t)}E_{k}^{(t)}+\delta^{(t)}\mathcal{E}_{k}^{(t)},b_{\max}\}.
 \end{align}
 From (\ref{up_cons_2}), in order to guarantee certain minimum data rate, we constrain the battery state of each sensor at the beginning of each time slot to be above a predefined threshold $b_0$, i.e.,
  \begin{align}\label{batt_constraint}
     b_k^{(t)}\geq b_0, k\in \mathcal{K}.
 \end{align}

\subsection{Outline of Results}
Note that the quantities \({\gamma}_{l,k}^{(t)}\), \({{\mathcal{E}}}_k^{(t)}\), and \({R}_k^{(t)}\) given in (\ref{Gamma_lk}), (\ref{received_sensor}), and (\ref{achievable_rate}) are functions of the random pilot sequences $\pmb{\Psi}$. In Section III, an asymptotic analysis is performed as \(\tau\rightarrow \infty\) while keeping \(\alpha\), \(\kappa={\tau}/K_a\) and \(E_p\) fixed, which shows that \({\gamma}_{l,k}\), \(\tilde{{\mathcal{E}}}_k(t)\), and \({R}_k(t)\) become independent of $\pmb{\Psi}$ in this regime, i.e., \begin{align}
    &\gamma_{l,k}^{(t)}-\bar{\gamma}_{l,k}^{(t)}\xrightarrow[]{\text{a.s.}}0, k\in \mathcal{K}, l=1,\cdots,L, \cr
    &\tilde{\mathcal{E}}_{k}^{(t)}-\bar{\mathcal{E}}_{k}^{(t)}\xrightarrow[]{\text{a.s.}}0, k\in \mathcal{K}, \cr
    &R_{k}^{(t)}-\bar{R}_{k}^{(t)}\xrightarrow[]{\text{a.s.}}0, k\in \mathcal{K},  \nonumber
\end{align}
where \(\tilde{\mathcal{E}}_{k}^{(t)}\) is a tight lower-bound of \({\mathcal{E}}_{k}^{(t)}\).
The closed-form expressions of \(\bar{\gamma}^{(t)}_{l,k}\), \(\bar{{\mathcal{E}}}^{(t)}_k\), and \(\bar{R}_k^{(t)}\) (that are independent of $\pmb{\Psi}$ ) can be used to predict the performance of the general case with large but finite length of pilot. 

In Section IV, we consider a long-term scheduling and power control problem to maximize the minimum time averaged achievable rate \(\min_{k\in \mathcal{K}}\mathop{\lim }_{T\rightarrow\infty}\frac{1}{T}\sum_{t=0}^{T-1}\bar{\delta}^{(t)}{R}_k^{(t)}\), while meeting all the constraints detailed in Section II.A. Specifically, given the large-scale fading coefficients $\{\beta_{l,k}\}$, the overall policy in time slot \(t\) consists of the transmission mode \(\delta^{(t)}\), sensor states \({\pmb \theta}^{(t)}\), downlink power control coefficients \({\pmb \eta}^{(t)}\) and uplink power control coefficients \({\pmb \xi }^{(t)}\), i.e., 
\begin{align}
    \mathcal{P}^{(t)}=\{{\delta^{(t)},} {\pmb \theta}^{(t)},{\pmb \eta }^{(t)}, {\pmb \xi }^{(t)}\}, \label{policy}
\end{align}
and the long-term optimization problem is formulated as
\begin{flalign}
{\bf P}_1: \max_{\{\mathcal{P}^{(t)}\}}~&\min_{k\in \mathcal{K}}\mathop{\lim }_{T\rightarrow\infty}\frac{1}{T}\sum_{t=0}^{T-1}{R}_k^{(t)}\label{P0.0}\\
{ s.t.} ~
&  (\ref{trans_mode}), (\ref{schedule_set}),(\ref{down_cons}), (\ref{AP_constraints}),  (\ref{up_cons_1}), (\ref{up_cons_2}), (\ref{bat_state}), (\ref{batt_constraint}). \nonumber
\end{flalign} 
In general, ${\bf P}_1$ is NP-hard due to the following reasons. Firstly, it is a max-min problem with an infinite horizon average objective function. Secondly, the constraint region is also over infinite horizon. In addition, it is a mixed integer program due to the 0-1 constraints w.r.t. the sensor states.

To overcome these difficulties, in Section IV, we first relax ${\bf P}_1$ into a maximization problem ${\bf P}_2$ with long-term constraints. Through establishing virtual queues, ${\bf P}_2$ is then reformulated into a maximization problem ${\bf P}'_2$ with the constraints on the rate stability of virtual queues. Using Lyapunov optimization, problem ${\bf P}'_2$ is then decomposed into a sequence of optimization problems of minimizing the Lyapunov drift plus penalty for each time slot. To solve the problem for each time slot, in Section V, we propose two low-complexity optimization methods for the downlink WPT and the uplink data transmission, respectively.

\section{Asymptotic Analysis as \(\tau\rightarrow \infty\)}

Based on (\ref{Z_l}), denote
\begin{align}
\pmb{Z}^{(t)}_{l,k}&=\pmb{Z}^{(t)}_l-E_p\beta_{l,k}{\pmb \psi}_k{\pmb \psi}_k^H\nonumber\\
&={\pmb I}_{\tau}+\sum\nolimits_{j\in \mathcal{K}_a^{(t)}/\{k\}}E_p\beta_{l,j}\pmb{\psi}_j\pmb{\psi}_j^H\nonumber\\
&={\pmb I}_{\tau}+\pmb{\Xi}_{l,k}\pmb{\Xi}_{l, k}^H,\nonumber\end{align}
where \({\pmb \Xi}_{ l,k}=\pmb{\Psi}_{k}\pmb{\Sigma}_{l,k}^{\frac1{2}}\), with \(\pmb{\Psi}_{k}=\left(\cdots,\pmb{\psi}_{j},\cdots\right)\), and \(\pmb{\Sigma}_{l,k}={\rm diag}\left(\cdots,E_p\beta_{l,j},\cdots\right)\) with $j\in \mathcal{K}_{a}^{(t)}/\{k\}$. Then, it is straightforward to obtain the following lemma according to Theorem 1 and Theorem 2 in \cite{J. Hoydis}. 

\begin{lemma}
As \(\tau\rightarrow \infty\), we have
\begin{align}\label{Lemma1.A}
{{\rm tr}\left[(\pmb{Z}_{l,k}^{(t)})^{-1}\right]}/{\tau}-\mathcal{Z}^{(t)}_{l,k}\xrightarrow[]{\text{a.s.}}0,
\end{align}
\begin{align}\label{Lemma1.B}
{{\rm tr}\left[(\pmb{Z}_{l,k}^{(t)})^{-2}\right]}/{\tau}-\tilde{\mathcal{Z}}^{(t)}_{l,k}\xrightarrow[]{\text{a.s.}}0.
\end{align}
  
\(\mathcal{Z}^{(t)}_{l,k}\) is given as
\[
\mathcal{Z}_{l,k}^{(t)}=\left[\sum\nolimits_{j\in \mathcal{K}_{a}^{(t)}/\{k\}}\frac{E_p\beta_{l,j}}{\tau(1+\varsigma_{j}) } +1 \right]^{-1},
\]
and \(\pmb{\varsigma}=[\cdots,\varsigma_{j},\cdots]^T,{j\in \mathcal{K}_{a}^{(t)}}/\{k\}\) is the unique solution to the following set of fixed-point equations
\[\varsigma_{j}={E_p\beta_{l,j}}\left[\sum\nolimits_{j\in \mathcal{K}_{a}^{(t)}/\{k\}}\frac{E_p\beta_{l,j}}{\tau(1+\varsigma_{j})} +1 \right]^{-1},\]
with initial values \(\varsigma_{j}=1\). \(\tilde{\mathcal{Z}}_{l,k}^{(t)}\) is given as
\begin{align}
  \tilde{\mathcal{Z}}_{l,k}^{(t)}=&\left[1+\frac{1}{\tau}\sum\nolimits_{j\in \mathcal{K}_{a}^{(t)}/\{k\}}\frac{E_p\beta_{l,j}\hat{\varsigma}_{j}}{(1+\varsigma_{j})^2 }\right]\left[{\mathcal{Z}}_{l,k}^{(t)}\right]^{-2},  \nonumber
\end{align}

and \(\hat{\pmb{\varsigma}}=\left[\cdots,\hat{\varsigma}_{j},\cdots\right]^T,{j\in \mathcal{K}_{a}^{(t)}/\{k\}}\) is given by
\begin{flalign}
&\hat{\pmb{\varsigma}}=({\pmb I}_{(K_a-1)}-{\pmb J})^{-1}{\pmb c},\nonumber\\
&[{\pmb J}]_{j,i}=\frac{E_p^2\beta_{l,j}\beta_{l,i}}{\tau(1+\varsigma_{j})^2 }\left[{\mathcal{Z}}_{l,k}^{(t)}\right]^{-2}, \nonumber\\
&[{\pmb c}]_j={E_p\beta_{l,j}}\left[{\mathcal{Z}}_{l,k}^{(t)}\right]^{-2}. \nonumber
\end{flalign}
\end{lemma}
\subsection{Asymptotic Analysis of \(\gamma_{l,k}^{(t)}\) }

\begin{theorem}
As \(\tau\rightarrow \infty\), the mean squared value of channel estimate given in (\ref{Gamma_lk}) almost surely converges to a deterministic value independent of the random pilots $\pmb{\Psi}$, i.e.,
     \begin{align}\label{gamma_app}
       {\gamma}_{l,k}^{(t)}-{\bar \gamma}_{l,k}^{(t)}\xrightarrow[]{\text{a.s.}}0, ~k\in \mathcal{K}_a^{(t)},
   \end{align}
   where 
   \begin{align}\label{bar_gamma}
   {\bar \gamma}_{l,k}^{(t)}=\frac{{E_p}\beta^2_{l,k}\mathcal{Z}_{l,k}^{(t)}}{1+{E_p}\beta_{l,k}\mathcal{Z}_{l,k}^{(t)}}.
   \end{align}
\end{theorem}
\begin{IEEEproof}
		See Appendix A.
\end{IEEEproof}

\subsection{Asymptotic Analysis of \(\mathcal{E}_{k}^{(t)}\) }
To analyze \(\mathcal{E}_{k}^{(t)}\) in (\ref{received_power}), we first introduce the following lemma from \cite{Xinhua Wang} on the LMMSE channel estimation with random pilots.
  \begin{lemma}\label{Sec3.1}
 	For \( m,l \in \left\{1,\cdots,L\right\}\) with \(m \neq l\) and \(k\in \mathcal{K}_{a}^{(t)}\), we have
 	\begin{align}
 	    &{\rm cov}\left[\hat{\pmb g}^{(t)}_{l,k},~\hat{\pmb g}^{(t)}_{m,k}\right]={\pmb 0},\nonumber\\
 	\mbox{and}\quad  &{\rm cov}\left[\left\|\hat{\pmb g}^{(t)}_{l,k}\right\|^2,~\left\|\hat{\pmb g}^{(t)}_{m,k}\right\|^2\right]={\pmb 0}.\nonumber
 	\end{align}
 \end{lemma}
 
 Lemma 2 reveals that the channel estimates between different APs and a sensor are uncorrelated. However, the channel estimates between an AP and different sensors are correlated, since \(\hat{g}^{(t)}_{(l,n),k}\) depends on \({g}^{(t)}_{(l,n),j}\) and \(\pmb{\psi}_{j}\) with \(j\in \mathcal{K}_{a}^{(t)}\), which can be seen from (\ref{received_pilots}) and (\ref{g_lnk_est}). 
Since the channel estimation error $\tilde{\pmb{g}}_{l,k}^{(t)}={\pmb{g}}_{l,k}^{(t)}-\hat{\pmb{g}}_{l,k}^{(t)}$ is independent of $\hat{\pmb{g}}_{l,k}^{(t)}$, $z_k^{(t,\ell)}$ in (\ref{received_sensor}) can be rewritten as 
	\begin{align}
	&z_k^{(t,\ell)}=\sum_{l=1}^L[\hat{\pmb{g}}^{(t)}_{l,k}]^T\pmb{x}_{l}^{(t,\ell)}+\sum_{l=1}^L[\tilde{\pmb{g}}^{(t)}_{l,k}]^T\pmb{x}_{l}^{(t,\ell)}+v_k^{(t,\ell)}\nonumber\\
	&=\underbrace{\sum_{l=1}^L\sqrt{\rho_d\eta_{l,k}^{(t)}\theta_k^{(t)}}[\hat{\pmb{g}}^{(t)}_{l,k}]^T[{\hat{\pmb{g}}^{(t)}_{l,k}}]^* q_k^{(t,\ell)}}_{\mathcal{S}_{k1}}+\nonumber\\
	&\underbrace{\sum_{l=1}^L\sum_{j\neq k}\sqrt{\rho_d\eta_{l,j}^{(t)}\theta_j^{(t)}}[\hat{\pmb{g}}^{(t)}_{l,k}]^T[{\hat{\pmb{g}}^{(t)}_{l,j}}]^* q_j^{(t,\ell)}+\sum_{l=1}^L[\tilde{\pmb{g}}^{(t)}_{l,k}]^T\pmb{x}_{l}^{(t,\ell)}+v_k^{(t,\ell)}}_{\mathcal{S}_{k2}}\nonumber
	\end{align}
Since \(\mathcal{S}_{k1}\) and \(\mathcal{S}_{k2}\) are uncorrelated and zero-mean, we have 
\begin{align}\label{Sec3.5}
{\mathcal{E}}_{k}^{(t)}&=(1-\alpha)\Delta \zeta\delta^{(t)}\theta_k^{(t)}\rho_d(T_c-\tau)\nonumber\\
&\qquad\qquad \times\mathbb{E}
\left[\left|\mathcal{S}_{k1}
\right|^2+\left|\mathcal{S}_{k2}\right|^2  +2\Re\{\mathcal{S}_{k1}\mathcal{S}_{k2}\}\right]\nonumber\\
 &\geq (1-\alpha)\Delta \zeta\delta^{(t)}\theta_k^{(t)}\rho_d(T_c-\tau)\mathbb{E}
\left[\left|\mathcal{S}_{k1}
\right|^2\right]\nonumber\\
 &= (1-\alpha)\Delta \zeta\delta^{(t)}\theta_k^{(t)}\rho_d\mathbb{E}\left[\left|\sum\nolimits_{l=1}^L\sqrt{\eta_{l,k}^{(t)}}[\hat{\pmb{g}}_{l,k}]^T[{\hat{\pmb{g}}^{(t)}_{l,k}}]^*\right|^2\right]\nonumber\\
 &= (1-\alpha)\Delta \zeta\delta^{(t)}\theta_k^{(t)}\rho_d\cr
 &\times\mathbb{E}\left[\sum\limits_{l=1}^L\sum\limits_{m=1}^L\sqrt{\eta_{l,k}^{(t)}\eta_{m,k}^{(t)}}[\hat{\pmb{g}}^{(t)}_{l,k}]^T[{\hat{\pmb{g}}^{(t)}_{l,k}}]^*[\hat{\pmb{g}}^{(t)}_{m,k}]^T[{\hat{\pmb{g}}^{(t)}_{m,k}}]^*\right]\nonumber\\
 &\overset{(a)}{=}(1-\alpha)\Delta \zeta\delta^{(t)}\theta_k^{(t)}\rho_d N\left[N\left(\sum\limits_{l=1}^{L}\sqrt{\eta_{l,k}^{(t)}}\gamma_{l,k}^{(t)}\right)^2\right.\nonumber\\
 &\left.+\sum\limits_{l=1}^{L}\eta_{l,k}^{(t)}[\gamma_{l,k}^{(t)}]^2\right]\nonumber\\
 &\geq \tilde{\mathcal{E}}_{k}^{(t)}\triangleq (1-\alpha)\Delta \zeta\delta^{(t)}\theta_k^{(t)}\rho_d N^2\left(\sum\limits_{l=1}^{L}\sqrt{\eta_{l,k}^{(t)}}\gamma_{l,k}^{(t)}\right)^2,
 \end{align}
where step (a) is obtained according to Lemma 2 and \(\mathbb{E}[||\hat{\pmb{g}}_{l,k}^{(t)}||^4]=N(N+1)[\gamma^{(t)}_{l,k}]^2\). It is straightforward to prove that $\tilde{\mathcal{E}}_{k}^{(t)}\rightarrow {\mathcal{E}}_{k}^{(t)}$ as $N\rightarrow \infty$, which implies $\tilde{\mathcal{E}}_{k}^{(t)}$ is a tight lower-bound on ${\mathcal{E}}_{k}^{(t)}$ for large $N$. Substituting (\ref{gamma_app}) into (\ref{Sec3.5}), we have 
 \begin{align}\label{Energy_Harvest_APP}
 &\tilde{\mathcal{E}}^{(t)}_{k}-\bar{\mathcal{E}}^{(t)}_{k}\xrightarrow[]{\text{a.s.}}0,~{\rm as}~\tau\rightarrow\infty, ~k\in \mathcal{K}_a^{(t)}
 \end{align}
 where
 \begin{align}
\bar{\mathcal{E}}^{(t)}_{k}&=(1-\alpha)\Delta \zeta\delta^{(t)}\theta_k^{(t)}\rho_d N^2\left(\sum\limits_{l=1}^{L}\sqrt{\eta_{l,k}^{(t)}}\bar{\gamma}_{l,k}^{(t)}\right)^2.\nonumber
\end{align}

\subsection{Asymptotic Analysis of \(R_{k}^{(t)}\) }
For given pilots $\pmb \Psi$, the SINR in (\ref{SINR}) can be expressed as \cite{Xinhua Wang}
    \begin{align}\label{Per_B.3}	
    \Gamma_k^{(t)}=\frac{\mathcal{D}_{k}^{(t)}\xi_k^{(t)}}{\mathcal{U}_{k}^{(t)}\xi_k^{(t)}+\sum_{j\in\mathcal{K}^{(t)}_{a}/\{k\}}\mathcal{I}^{(t)}_{k,j}\xi^{(t)}_j+\mathcal{N}^{(t)}_{k}
	},
    \end{align}
   where
   \begin{align}
    \mathcal{D}_{k}^{(t)}&=\rho_uN\left(\sum\nolimits_{l=1}^L\gamma_{l,k}^{(t)} \right)^2,
    \mathcal{U}_{k}^{(t)}=\sum\nolimits_{l=1}^L \rho_u\gamma_{l,k}^{(t)}\beta_{l,k},\nonumber \\
    \mathcal{N}^{(t)}_{k}&=\sigma^2\sum\nolimits_{l=1}^L\gamma_{l,k}^{(t)}, \mbox{and} \nonumber\\
   \mathcal{I}^{(t)}_{k,j}&=\rho_u\sum\nolimits_{l=1}^L\beta_{l,j}\left\|\pmb{a}_{l,k}^{(t)}\right\|^2+\rho_uE_p N\Big|\sum\nolimits_{l=1}^L\beta_{l,j}\pmb{\psi}_j^H\pmb{a}^{(t)}_{l,k}
   \Big|^2\nonumber \\
    &+\rho_uE_p\sum\nolimits_{l=1}^L\sum\nolimits_{i=1}^{K}\beta_{l,j}\beta_{l,i}|\pmb{\psi}_i^H\pmb{a}^{(t)}_{l,k}|^2.\nonumber
   \end{align}
 
	\begin{theorem}\label{Sec3.10}
	As \(\tau\rightarrow \infty\), the rate in (\ref{achievable_rate}) almost surely converges to a deterministic value independent of the random pilots ${\pmb \Psi}$, i.e.,
	\begin{align}
	    &{R}_{k}^{(t)}-\bar{ R}_{k}^{(t)}\xrightarrow[]{\text{a.s.}}0, ~k\in \mathcal{K}_a^{(t)},\nonumber
\end{align}
where
\begin{align}\label{Achievablerates_APP}
	    &\bar{R}_{k}^{(t)}=
	(1-\alpha)\bar{\delta}^{(t)} \theta_k^{(t)}\log_2
	\left(
	1 + \bar{\Gamma}_k^{(t)}
	\!\right), {\rm ~~bits/s/Hz},
	\end{align}
	\begin{align}
	 \bar{\Gamma}^{(t)}_k=\frac{\bar{D}^{(t)}_{k}\xi_k^{(t)}}{\bar{U}^{(t)}_{k}\xi_k^{(t)}+\sum_{j\in\mathcal{K}^{(t)}_{a}/\{k\}}\bar{\mathcal{I}}^{(t)}_{k,j}\xi_{j}^{(t)}+\bar{\mathcal{N}}^{(t)}_{k}
	},\nonumber
	\end{align}
\(\bar{D}_{k}^{(t)}= N\rho_u(\sum\nolimits_{l=1}^L\bar{\gamma}_{l,k}^{(t)} )^2\), \(\bar{U}^{(t)}_{k}=\sum\nolimits_{l=1}^L \rho_u\bar{\gamma}^{(t)}_{l,k}\beta_{l,k}\),  \(\bar{\mathcal{I}}^{(t)}_{k,j}=\rho_u\sum\nolimits_{l=1}^L\beta_{l,j}\varrho^{(t)}_{l,k}+\rho_u E_p\sum\nolimits_{l=1}^L\beta_{l,j}\beta_{l,k}\vartheta^{(t)}_{l,k},\) and \(\bar{\mathcal{N}}^{(t)}_{k}=\sigma^2\sum\nolimits_{l=1}^L\bar{\gamma}^{(t)}_{l,k}\)
    with
    \begin{align}
    \varrho_{l,k}^{(t)}&=\frac{E_p\beta_{l,k}^2\tilde{\mathcal{Z}}^{(t)}_{l,k}}{\left(1+E_p\beta_{l,k}{\mathcal{Z}}^{(t)}_{l,k}\right)^2}, \nonumber\\
    \vartheta_{l,k}^{(t)}&=\frac{E_p\beta_{l,k}^2[{\mathcal{Z}}^{(t)}_{l,k}]^2}{\left(1+E_p\beta_{l,k}{\mathcal{Z}}^{(t)}_{l,k}\right)^2}.\nonumber 
    \end{align}
	\end{theorem}
	\begin{IEEEproof}
	See Appendix B.
	\end{IEEEproof}
	
\subsection{Impact of More Active Sensors}
It is noted that ${\bar \gamma}_{l,k}^{(t)}$, $\bar{\mathcal{E}}^{(t)}_{k}$, and $\bar{R}_{k}^{(t)}$ are functions of the large-scale fading coefficients of active sensors. We now examine the effect of adding more sensor to the current set $\mathcal{K}_a^{(t)}$ on each sensor $k\in \mathcal{K}_a^{(t)}$. In particular, let \(\bar{\gamma}_{l,k}^{(t)}\) be the asymptotic mean squared value of channel estimate given in (\ref{bar_gamma}), and \([\bar{\gamma}^{(t)}_{l,k}]'\) be the corresponding value when the active set becomes \([\mathcal{K}_{a}^{(t)}]'\supset \mathcal{K}_{a}^{(t)}\) by activating some inactive sensors in $\mathcal{K}/\mathcal{K}_{s}^{(t)}$. We have the following result:
\begin{lemma}
\begin{align}
 [\bar{\gamma}^{(t)}_{l,k}]'<\bar{\gamma}_{l,k}^{(t)}, k\in \mathcal{K}_a^{(t)}, l=1,\cdots,L.
\end{align}
\end{lemma}
\begin{IEEEproof}
		See Appendix C.
\end{IEEEproof}

\emph{\textbf{Remark 1}}: The mean squared error of the channel estimate is 
	\begin{align}\label{mean_error}
    e_{l,k}^{(t)}=\mathbb{E}\left[\left\|{g}_{(l,n),k}^{(t)}-{\hat g}_{(l,n),k}^{(t)}\right\|^2\right]=\beta_{l,k}-\gamma_{l,k}^{(t)}.
\end{align}
Lemma 3 reveals that the accuracy of channel estimation is reduced by activating inactive sensors. The reduced $\bar{\gamma}_{l,k}^{(t)}$ further degrades the amount of harvested energy $\bar{\mathcal{E}}_{k}^{(t)}$ in (\ref{Energy_Harvest_APP}) and the achievable rate $\bar{R}_{k}^{(t)}$ in (\ref{Achievablerates_APP}), since they are both monotonically increasing functions w.r.t. \(\bar{\gamma}_{l,k}^{(t)}\). Specifically, the influence on the efficiency of WPT is more pronounced, since it is proportional to the square of \(\bar{\gamma}_{l,k}^{(t)}\).

\section{Long-term Scheduling and Power Control }

In ${\bf P}_1$, we aim to achieve the max-min  fairness over infinite horizon, which is hard to tackle. Similarly as in \cite{Z. Jing}, we first introduce a sequence of auxiliary variables $r^{(t)}$ bounded between $0$ and $r_{\max}$, where $r_{\max}=\mathop{\max}\limits_{k\in \mathcal{K}}\hat{R}_k^{(t)}$, with $$\hat{R}_k^{(t)}=(1-\alpha) \theta_k^{(t)}\log_2
	\left(
	1 + \frac{N\rho_u}{\sigma^2}\sum\nolimits_{l=1}^L{\beta}_{l,k} \right)
,$$ i.e., $\hat{R}_k^{(t)}$ is the achievable rate of the $k$-th sensor in (\ref{Achievablerates_APP}) with full transmit power, perfect channel estimation and no interference. Then, we have 
\begin{align}
    0\leq \max_{k\in \mathcal{K}}R_k^{(t)}\leq r_{\max}, \nonumber
\end{align}
for any $t$, which implies
\begin{align}
    0\leq \min_{k\in \mathcal{K}}\mathop{\lim }_{T\rightarrow\infty}\frac{1}{T}\sum_{t=0}^{T-1} R_k^{(t)}\leq r_{\max}.\nonumber
\end{align}
Thus, the max-min problem ${\bf P}_1$ is equivalent to maximizing $\mathop{\lim }_{T\rightarrow\infty}\frac{1}{T}\sum_{t=0}^{T-1}r^{(t)}$ with the following two extra constraints \cite{Z. Jing, M. J. Neely},
\begin{align}
    & \mathop{\lim \inf}_{T\rightarrow\infty}\frac{1}{T}\sum_{t=0}^{T-1}[{R}_k^{(t)}-r^{(t)}]\geq 0,k\in \mathcal{K},  \label{P2.1}\\
& 0\leq r^{(t)}\leq r_{\max}.\label{P2.2}
\end{align}
Another challenge resides in constraint (\ref{batt_constraint}), which makes the control policy $\mathcal{P}^{(t)}$ over different time slots coupled due to the dynamics of $b_k^{(t)}$ in (\ref{bat_state}). We therefore relax (\ref{batt_constraint}) to the following long-term constraint, 
\begin{align}
    \mathop{\lim \inf}_{T\rightarrow\infty} \frac{1}{T}\sum_{t=1}^{T-1}b_j^{(t)}\geq b_0, k\in \mathcal{K},\label{long_term_cons}
\end{align}
which enable us to design the policy using Lyapunov optimization.
Hence, ${\bf P}_1$ is relaxed to 
\begin{flalign}
{\bf P}_2: \max_{\{\mathcal{P}^{(t)},r^{(t)}\}}~&\mathop{\lim }_{T\rightarrow\infty}\frac{1}{T}\sum_{t=0}^{T-1}r^{(t)} \nonumber\\
{ s.t.} ~& (\ref{P2.1}),(\ref{P2.2}), (\ref{long_term_cons}),\nonumber \\
&  (\ref{trans_mode}), (\ref{schedule_set}), (\ref{down_cons}), (\ref{AP_constraints}), (\ref{up_cons_1}), (\ref{up_cons_2}), (\ref{bat_state}). \nonumber
\end{flalign} 

\subsection{Problem Reformulation with Queue Stability Constraints} 
${\bf P}_2$ is still hard to tackle due to the long-term constraints (\ref{P2.1}) and (\ref{long_term_cons}). Similarly as in \cite{D. Zhai,M. Peng}, we transform the long-term constraints into queue stability constraints.

Define \(\{X_{k}(t):k\in \mathcal{K}\}\) as the virtual queues associated with constraint (\ref{long_term_cons}). In each time slot, the virtual queue \(X_{k}(t)\) is updated according to
\begin{align}\label{backlog_x}
X_{k}(t+1)=\left[X_{k}(t)+b_0-b_{k}^{(t+1)}\right]^{+},
\end{align}
where \([x]^{+}=\max\{0,~x\}\). $b_0$ and $b_{k}^{(t+1)}$ can be considered as the arrival rate and the departure rate of the virtual queue \(X_{k}(t)\), respectively. We say \(X_k(t)\) is rate stable if \(\lim\limits_{t\rightarrow \infty}\frac{X_k(t) }{t}=0 \) \cite{D. Zhai,M. Peng}. To maintain the rate stability of \(X_{k}(t)\), the departure rate $b_{k}^{(t+1)}$ must be no less than the arrival rate $b_0$, which coincides with constraint (\ref{long_term_cons}). 

Similarly, we also define \(\{Y_{k}(t):k\in \mathcal{K}\}\) as the virtual queues associated with constraint (\ref{P2.1}), where \(Y_{k}(t)\) is updated as
 \begin{align}\label{backlog_y}
Y_{k}(t+1)=\left[Y_{k}(t)+r^{(t)}-{R}_{k}^{(t)}\right]^{+}.
\end{align}
To reveal the relation between the long-term constraints (\ref{P2.1}) and (\ref{long_term_cons}) in ${\bf P}_2$ and the rate stability of $\{{X}_{k}(t),Y_{k}(t):k\in \mathcal{K}\}$, we have the following lemma.
\begin{lemma}\label{Sec4.4}
If $\{{X}_{k}(t),Y_{k}(t):k\in \mathcal{K}\}$ are rate stable with finite initial values, then the long-term constraints (\ref{P2.1}) and (\ref{long_term_cons}) are satisfied.
\end{lemma}
\begin{IEEEproof}
Without loss of generality,	we take \(X_{k}(t)\) for example. (\ref{backlog_x}) can be rewritten as
	\begin{align}
		X_{k}(t+1)=X_{k}(t)-b_{k}^{(t+1)}
		+\max\left\{b_0,b_{k}^{(t+1)}-X_{k}(t)\right\}.\nonumber
	\end{align}
	Sum over $t=0,\cdots, T-1$, we have 
	\begin{align}\label{Sec4.6}
			X_{k}(T)-X_{k}(0)&=\sum_{t=0}^{T-1}\left\{\max\left\{b_0,b_{k}^{(t+1)}-X_{k}(t)\right\}- b_{k}^{(t+1)}\right\}\nonumber\\
			&\geq T b_0- \sum_{t=1}^{T}b_{k}^{(t)}.
	\end{align}
	Dividing both sides of (\ref{Sec4.6}) by $T$ and taking the limit, we obtain
		\begin{align}\label{Sec4.7}
		\mathop{\lim \inf}\limits_{T\rightarrow \infty}\frac{X_{k}(T)}{T}&-\mathop{\lim \inf}\limits_{T\rightarrow \infty}\frac{X_{k}(0)}{T}\geq b_0- \mathop{\lim \inf}\limits_{T\rightarrow \infty}\frac{1}{T}\sum_{t=1}^{T}b_{k}^{(t)}\nonumber\\
		&\overset{(a)}{=}b_0- \mathop{\lim \inf}\limits_{T\rightarrow \infty}\frac{1}{T}\sum_{t=0}^{T-1}b_{k}^{(t)},
		\end{align}
		where (a) is due to $0\leq b_{k}^{(t)}\leq b_{\max}$. Hence, the rate stability of $X_{k}(t)$ and the finiteness of $X_{k}(0)$ imply \(\mathop{\lim \inf} \frac{1}{T}\sum_{t=1}^{T-1}b_{k}^{(t)}\geq b_0\). 
\end{IEEEproof}
Based on Lemma 3, we then replace the long-term constraints (\ref{P2.1}) and (\ref{long_term_cons}) in ${\bf P}_2$ by rate stability constraints, leading to the following formulation with stricter constraints. 
\begin{flalign}
{\bf P'}_2: \max_{\{\mathcal{P}^{(t)}, r^{(t)}\}}~&\mathop{\lim }_{T\rightarrow\infty}\frac{1}{T}\sum_{t=0}^{T-1}r^{(t)} \nonumber\\
{ s.t.} ~& \{{X}_{k}(t),Y_{k}(t):k\in \mathcal{K}\}\mbox{ are rate stable,} \nonumber\\
&  (\ref{trans_mode}), (\ref{schedule_set}), (\ref{down_cons}),(\ref{AP_constraints}), (\ref{up_cons_1}), (\ref{up_cons_2}), (\ref{bat_state}),(\ref{P2.2}). \nonumber
\end{flalign} 

\subsection{Lyaopunov Drift Plus Penalty Method}
To investigate the rate stability of all virtual queues $\{{X}_{k}(t),Y_{k}(t):k\in \mathcal{K}\}$, we define the Lyapunov function \cite{M. J. Neely}
\begin{align}\label{Sec4.10}
\mathcal{L}(t)=\sum\limits_{k=1}^K\frac{X^2_k(t)}{2}+\sum\limits_{k=1}^K\frac{Y^2_k(t)}{2}.
\end{align}
The Lyaopunov drift is defined as
\begin{align}\label{drift_def}
    \mathcal{D}(t)=\mathcal{L}(t+1)-\mathcal{L}(t).
\end{align}
Let $[\mathcal{P}^{(t)}]^*$ denote the policy that minimizes the Lyaopunov drift in each time slot, i.e.,
\begin{align}\label{Sec4.11a}
[\mathcal{P}^{(t)}]^*=\mathop{\arg \min}_{\mathcal{P}{(t)}}~     \mathcal{D}(t), t=0,1,\cdots
\end{align}
Then, we have the following theorem.
 \begin{theorem}\label{Sec4.12}
  Let $\{X^*_k(t),Y^*_k(t):k\in \mathcal{K}\}$ denote the virtual queues corresponding to policy \([\mathcal{P}^{(t)}]^*\) defined in (\ref{Sec4.11a}), while  $\{X'_k(t),Y'_k(t):k\in \mathcal{K}\}$ denote the virtual queues of any other feasible policy \([\mathcal{P}^{(t)}]'\) different from \([\mathcal{P}^{(t)}]^*\). If $\{X'_k(t),Y'_k(t):k\in \mathcal{K}\}$ are rate stable, then $\{X^*_k(t),Y^*_k(t):k\in \mathcal{K}\}$ must be rate stable.
 \end{theorem}
\begin{IEEEproof}
See Appendix D.
\end{IEEEproof}
Theorem 3 reveals that the minimizer of \(\mathcal{D}(t)\) in each time slot is more likely to achieve the rate stability of $\{{X}_{k}(t),Y_{k}(t):k\in \mathcal{K}\}$ compared with any other feasible policies. In ${\bf P}'_2$, we aim to maximize $\mathop{\lim }_{T\rightarrow\infty}\frac{1}{T}\sum_{t=0}^{T-1}r^{(t)}$ while keeping all queues $\{{X}_{k}(t),Y_{k}(t):k\in \mathcal{K}\}$ rate stable. To achieve this goal, we minimize the following drift plus penalty in each time slot \cite{M. J. Neely}, i.e.,
\begin{align}\label{Drift_plus_Panelty}
  \mathop{\min}_{\mathcal{P}^{(t)},r^{(t)}} \varPhi\left(t\right)=\mathcal{D}(t)-Wr^{(t)}, t=1,2,\cdots,T,
\end{align}
where \(W\) is a positive weight to balance the two items in (\ref{Drift_plus_Panelty}). However, minimizing \(\varPhi\left(t\right)\) is not easy to tackle due to the quadratic form of the Lyapunov drift $\mathcal{D}(t)$. Similarly as in \cite{M. J. Neely}, we will minimize the upper-bound \(\bar{\varPhi}\left(t\right)\) given in the following theorem instead of minimizing \(\varPhi\left(t\right)\) directly. 
\begin{theorem}
\(\varPhi\left(t\right)\) is upper bounded by the following linear function
\begin{align}
\bar{\varPhi}\left(t\right)&=\sum_{k=1}^K\{X_{k}(t)[b_0-b_{k}^{(t+1)}]+{C}_{0}\nonumber\\
&+Y_{k}(t)[r^{(t)}-{R}_k^{(t)}]+\bar{C}_{0}\}-Wr^{(t)}\nonumber,
\end{align}
where \({C}_{0}=b_{\max}^2/2\) and \(\bar{C}_{0}=r_{\max}^2/2\). 
\end{theorem}
\begin{IEEEproof}
According to (\ref{backlog_x}), we have 
\begin{align}\label{APP_E.1}
&X_{k}(t+1)^2-X_{k}(t)^2\nonumber\\
&=\left(\left[X_{k}(t)-b_0+b_{k}^{(t+1)}\right]^{+}\right)^2-X_{k}(t)^2\nonumber\\
&\leq\left[X_{k}(t)-b_0+b_{k}^{(t+1)}\right]^2-X_{k}(t)^2\nonumber\\
&=\left[b_0-b_{k}^{(t+1)}\right]^2+2X_{k}(t)\left[b_0-b_{k}^{(t+1)}\right]\nonumber\\
&\leq b_{\max}^2+2X_{k}(t)\left[b_0-b_{k}^{(t+1)}\right].
\end{align}

From (\ref{backlog_x}), we have 
\begin{align}\label{APP_E.3}
&Y_{k}(t+1)^2-Y_{k}(t)^2\nonumber\\
&=\left(\left[Y_{k}(t)-r^{(t)}+R_{k}^{(t)}\right]^{+}\right)^2-Y_{k}(t)^2\nonumber\\
&\leq\left[Y_{k}(t)-r^{(t)}+R_{k}^{(t)}\right]^2-Y_{k}(t)^2\nonumber\\
&=\left[r(t)-R_{k}^{(t)}\right]^2+2Y_{k}(t)\left[r^{(t)}+R_{k}^{(t)}\right]\nonumber\\
&\leq r_{\max}^2+2Y_{k}(t)\left[r^{(t)}+R_{k}^{(t)}\right]
\end{align}
Substituting (\ref{APP_E.1}) and (\ref{APP_E.3}) into (\ref{Drift_plus_Panelty}), we can conclude the proof.
\end{IEEEproof}
It is noted that minimizing the upper-bound $\bar{\varPhi}\left(t\right)$ is also helpful to strike a balance between the objective function and the rate stability of virtual queues. For example, if $X_k(t)$ tends to rate unstable, then we have $X_k(t)\rightarrow \infty$ and $X_k(t)+b_0-b_{j}^{(t+1)}\gg 0$. According to (\ref{drift_def}) and (\ref{APP_E.1}), we have $X_{k}(t)\left[b_0-b_{k}^{(t+1)}\right]\rightarrow \mathcal{D}(t)$, which implies minimizing $\bar{\phi}(t)$ is equivalent to minimizing ${\phi}(t)$. Since the backlogs \(\{{X}_k(t),{Y}_k(t):k\in \mathcal{K}\}\) are known at time slot $t$, minimizing $\bar{\varPhi}\left(t\right)$ is equivalent to maximizing 
\begin{align}\label{def_Drift_pen}
 \tilde{\varPhi}\left(t\right)&=\sum\nolimits_{k=1 }^K[X_{k}(t)b_{k}^{(t+1)}+Y_{k}(t){R}_k^{(t)}]\cr
    &+[W-\sum\nolimits_{k=1 }^KY_{k}(t)]r^{(t)}.
\end{align}
We can interpret the maximization of $\tilde{\varPhi}\left(t\right)$ as follows. The violation of the constraint $b_{k}^{(t+1)}>b_0$ leads to the growth of the backlog \(X_{k}(t)\). When \(X_{k}(t)\) is sufficiently large to dominate $\tilde{\varPhi}\left(t\right)$, maximizing $\tilde{\varPhi}\left(t\right)$ is equivalent to maximizing $b_{k}^{(t+1)}$, which tends to satisfy $b_{k}^{(t+1)}>b_0$ and further reduce \(X_{k}(t+1)\). Similarly, a large backlog \(Y_{k}(t)\) leads to maximizing ${R}_k^{(t)}$ to reduce \(Y_{k}(t+1)\).

\section{Optimization methods for each time slot}
In this section, we determine the optimal policy $$\mathcal{P}^{(t)}=\{{\delta^{(t)},} {\pmb \theta}^{(t)},{\pmb \eta }^{(t)}, {\pmb \xi }^{(t)}\}$$ for each time slot $t=0,1,2,\cdots,T$, by solving 
\begin{align}
{\bf P}_3:\max_{\mathcal{P}^{(t)}, r^{(t)}} &\quad  \tilde{\varPhi}\left(t\right)=\sum\nolimits_{k=1 }^K[X_{k}(t)b_{k}^{(t+1)}+Y_{k}(t){R}_k^{(t)}]\cr
    &+(W-\sum\nolimits_{k=1 }^KY_{k}(t))r^{(t)}\nonumber\\
{ s.t.} &   (\ref{trans_mode}), (\ref{schedule_set}), (\ref{down_cons}),(\ref{AP_constraints}), (\ref{up_cons_1}), (\ref{up_cons_2}), (\ref{bat_state}),(\ref{P2.2}). \nonumber
\label{Sec5.1}
\end{align} 
 In ${\bf P}_3$, $r^{(t)}$ is independent of $\mathcal{P}^{(t)}$, hence the optimal $r^{(t)}$ depends only on the backlogs \(\{Y_k(t): k\in \mathcal{K}\}\), i.e.,   
 \begin{equation}
\label{r_optimal}
[r^{(t)}]^*=\left\lbrace 
\begin{aligned}
r_{\max},&\qquad {\rm if}~\sum\nolimits_{k=1 }^KY_{k}(t)\leq W,\\
0, &\qquad  {\rm otherwise}.\\
\end{aligned} 
 \right. 
\end{equation}
Next, given $\delta^{(t)}$, we set the value of \(\pmb{\theta}^{(t)}\), or equivalently the set of active sensors $\mathcal{K}_a^{(t)}$, in a greedy way. Specifically, we choose the $K_a$ active sensors with the largest \(X_{k}(t)\) for $\delta^{(t)}=1$ and choose the $K_a$ active sensors with the largest \(Y_{k}(t)\) for $\delta^{(t)}=0$. Then ${\bf P}_3$ is decomposed into the following two sub-problems. For $\delta^{(t)}=1$, we aim to optimize the WPT through the downlink power control coefficients \({\pmb \eta}^{(t)}\), i.e.,
\begin{align}
{\bf P}_{3}^{'}:\max_{{\pmb \eta}^{(t)}} &\quad \sum\nolimits_{k\in \mathcal{K}_a^{(t)} }X_{k}(t)b_{k}^{(t+1)}\nonumber\\
{ s.t.} &\quad  (\ref{down_cons}),(\ref{AP_constraints}), (\ref{bat_state}). \nonumber
\end{align}
And for \(\delta^{(t)}=0\), we aim to optimize the data transmission through the uplink power control coefficients \({\pmb \xi}^{(t)}\), i.e.,
\begin{align}
{\bf P}_{3}^{''}:\max_{ {\pmb{ \xi}^{(t)}}} &~ \sum\nolimits_{k \in \mathcal{K}_a^{(t)} }[X_{k}(t)b_{k}^{(t+1)}+Y_{k}(t){R}_k^{(t+1)}]\nonumber\\
{ s.t.}&~  (\ref{up_cons_1}), (\ref{up_cons_2}), (\ref{bat_state}).\nonumber
\end{align}

Denote \(Q'\) and \(Q''\) as the optimal objective values of ${\bf P}_{3}^{'}$ and ${\bf P}_{3}^{''}$, respectively. Then, the transmission mode \(\delta^{(t)}\) is determined as
\begin{equation}
\label{mode_selection}
[\delta^{(t)}]^*=\left\lbrace 
\begin{aligned}
1,&\qquad {\rm if}~Q'\geq Q'',\\
0, &\qquad  {\rm otherwise}.\\
\end{aligned} 
 \right. 
\end{equation}

\subsection{Solution to ${\bf P}_{3}^{'}$}
 Given \(\delta^{(t)}=1\), the battery state is updated according to
 \begin{align}
 b_k^{(t+1)}=\min\{b_k^{(t)}+\tilde{\mathcal{E}}_{k}^{(t)},b_{\max}\}\nonumber.
 \end{align}
Thus, ${\bf P}_{3}^{'}$ becomes
\begin{align}
\max_{{\pmb \eta}^{(t)}} &\quad \sum\limits_{k\in \mathcal{K}^{(t)}_a }X_{k}(t)\min[\tilde{\mathcal{E}}_k(t),b_{\max}-b_k^{(t)}]\label{P3a.1}\\
{ s.t.} &\quad  (\ref{down_cons}),(\ref{AP_constraints}), \nonumber
\end{align}
which is NP-hard due to the non-convex objective function. The difference $b_{\max}-b_k^{(t)}$ is usually much greater than $\tilde{\mathcal{E}}_k(t)$ since $b_k^{(t)}$ always fluctuates around $b_0$ which can be seen from the simulation results. Thus, the objective function (\ref{P3a.1}) becomes
\begin{align}\label{P3a.2}
   f({\pmb \mu})=\sum\nolimits_{k\in \mathcal{K}^{(t)}_a }X_{k}(t)\tilde{\mathcal{E}}_k(t),
\end{align}
where $\tilde{\mathcal{E}}_k(t)$ given in (\ref{Sec3.5}) is a function of  $\mu^{(t)}_{l,k}=\sqrt{{\eta}^{(t)}_{l,k}}{\gamma}^{(t)}_{l,k}$. Using the sequential convex programming method \cite{S. Boyd, S. K. Joshi}, we can find a solution to (\ref{P3a.1}) by sequentially maximizing the first order approximation of $f({\pmb \mu})$ 
$$\widehat{f}({\pmb \mu})={f}(\widehat{\pmb \mu})+\sum_{l=1}^{L}\sum_{k\in \mathcal{K}^{(t)}_a}\frac{\partial {f}(\widehat{\pmb \mu})}{\partial \widehat{\mu}_{l,k}}(\mu_{l,k}-\widehat{\mu}_{l,k}),$$
near a feasible point $\widehat{\pmb \mu}$ which is updated after each iteration, i.e.,
\begin{align}
\max_{{\pmb \mu}} &\quad \widehat{f}({\pmb \mu})\label{P3a.3}\\
{ s.t.} &\quad  {\pmb \mu}\in \mathcal{T},\nonumber\\
&\quad {\pmb \mu}\geq 0,\nonumber\\
&\quad \sum\nolimits_{k\in \mathcal{K}^{(t)}_a}{[\mu^{(t)}_{l,k}]^2}/{\gamma^{(t)}_{l,k}}\leq 1, \forall l,\label{P3a.4}
\end{align}
where $ \mathcal{T}=\{{\pmb \mu}: |{\pmb \mu}-\widehat{\pmb \mu}|\leq \rho_0\}$ is the trust region around point $\widehat{\pmb \mu}$, and (\ref{P3a.4}) is the maximum transmit power constraint of each AP resulting from (\ref{AP_constraints}). After obtaining the optimal solution ${\pmb \mu}^{*}$ of the convex problem (\ref{P3a.3}), we update $\widehat{\pmb \mu}={\pmb \mu}^{*}$. Until convergence, the WPT power control coefficients ${\pmb \eta}^{(t)}$ is determined by 
$[{\eta}^{(t)}_{l,k}]^*=({[\mu^{(t)}_{l,k}]^*}/{\gamma^{(t)}_{l,k}})^2$.

\subsection{Solution to ${\bf P}_{3}^{''}$ }

 Given \(\delta^{(t)}=0\), the battery state is updated according to \[b_{k}^{(t+1)}=b_{k}^{(t)}-(1-\alpha) \rho_u \theta_j^{(t)}\xi_j^{(t)}.\] Hence, ${\bf P}_{3}^{''}$ can be rewritten as
\begin{align}
\max_{{\pmb \xi}^{(t)}} &\quad \sum\nolimits_{k\in \mathcal{K}^{(t)}_a} [Y_{k}(t){R}_k^{(t)}-(1-\alpha) X_{k}(t)\rho_u\xi_k^{(t)}]\label{P3B.1}\\
{ s.t.} & \quad  (\ref{up_cons_1}), (\ref{up_cons_2}), \nonumber
\end{align}
which aim to strike a balance between the weighted sum rate and the weighted transmit power consumption. Solving (\ref{P3B.1}) is equivalent to solving a series of sub-problems for fixed $$\sum\nolimits_{k\in \mathcal{K}^{(t)}_a}(1-\alpha) X_{k}(t)\rho_u\xi_k^{(t)}= \chi$$ with $\chi \in \left(0,(1-\alpha) \sum\nolimits_{k\in \mathcal{K}^{(t)}_a}X_{k}(t)\rho_u\right)$. According to (\ref{achievable_rate}) and (\ref{Per_B.3}), the sub-problems can be written as
 \begin{align}
\max_{{\pmb \xi}^{(t)}} &\quad \sum\nolimits_{k\in \mathcal{K}^{(t)}_a} (1-\alpha)Y_{k}(t)\log_2\left(1+\frac{A_{k}^{(t)}}{B_{k}^{(t)}}\right) \label{P3B.2}\\
{ s.t.} & \quad \sum\nolimits_{k\in \mathcal{K}^{(t)}_a}(1-\alpha) X_{k}(t)\rho_u\xi_k^{(t)}= \chi, \label{P3B.2a}\\
&\quad (\ref{up_cons_1}), (\ref{up_cons_2}), \nonumber
\end{align}
where $$A_{k}^{(t)}=\mathcal{D}_{k}^{(t)}\xi_k^{(t)}$$ and $$B_{k}^{(t)}=\mathcal{U}_{k}^{(t)}\xi_k^{(t)}+\sum\nolimits_{j\in\mathcal{K}^{(t)}_{a}/\{k\}}\mathcal{I}^{(t)}_{k,j}\xi^{(t)}_j+\mathcal{N}^{(t)}_{k},$$
which implies (\ref{P3B.2}) is a standard weighted sum-of-logarithms
maximization problem \cite{H. Guo, K. Shen_A,K. Shen_B}. Introducing the auxiliary variables $\{\omega_k, k\in \mathcal{K}^{(t)}_a\}$, 
(\ref{P3B.2}) is equivalent to  
\begin{align}
\max_{{\pmb \xi}^{(t)}, \{\omega_k\}} &\quad \sum\limits_{k\in \mathcal{K}^{(t)}_a}(1-\alpha) Y_{k}(t)\left[\log_2(1+\omega_k)-\omega_k\right]\nonumber\\
&+\sum\limits_{k\in \mathcal{K}^{(t)}_a}\frac{(1-\alpha)Y_{k}(t)(1+\omega_k)A_k^{(t)}}{A_k^{(t)}+B_k^{(t)}}, \label{P3B.3}\\
{ s.t.}  &\quad (\ref{up_cons_1}), (\ref{up_cons_2}), (\ref{P3B.2a}), \nonumber
\end{align}
which has been proved in \cite{K. Shen_B}. Then, we alternately solve ${\pmb \xi}^{(t)}$ and $\{\omega_k\}$ while fixing the other. For fixed ${\pmb \xi}^{(t)}$, the optimal $\{\omega_k\}$ is given by $\omega_k^*=\Gamma_k^{(t)}$ using the KKT condition. For fixed $\{\omega_k\}$, (\ref{P3B.3}) is reduced to a sum-of-ratios maximization problem which can be solved using the quadratic transform based fractional optimization technique proposed in \cite{K. Shen_A,K. Shen_B}.
\subsection{Overall Algorithm}
Finally, the proposed long-term scheduling and power control algorithm for solving ${\bf P}_1$ in (\ref{P0.0}) is summarized in Algorithm 1.
\begin{algorithm}
\caption{Proposed long-term scheduling and power control algorithm.}
\begin{algorithmic}[1] \label{Alg:GAP_algorithm}
\STATE {\bf Input}: The large-scale fading coefficients $\{\beta_{l,k},l=1,\cdots,L,~ k\in \mathcal{K}\}.$
\STATE {\bf Initialization}: Set \(T_{\max}\) and initialize the backlogs \(\{{X}_k(0)={Y}_k(0)=0:k\in \mathcal{K}\}\).
\FOR{$t=1:T_{\max}$}
\STATE According to the current backlogs \(\{{Y}_k(t):k\in \mathcal{K}\}\), obtain the optimal auxiliary variables $[r^{(t)}]^*$ according to (\ref{r_optimal}).
\STATE Solve problem ${\bf P}_{3}^{'}$ according to Section V.A, obtain the optimal solution $[{\pmb \eta}^{(t)}]^*$ and the corresponding objective value $Q'$.
\STATE Solve problem ${\bf P}_{3b}$ according to Section V.B, obtain the optimal solution $[{\pmb \xi}^{(t)}]^*$ and the corresponding objective value $Q''$.
\STATE Determine the optimal transmission mode $[\delta^{(t)}]^*$ and the corresponding sensor states $[{\pmb \theta}^{(t)}]^*$ according to (\ref{mode_selection}). and select the optimal corresponding power control coefficients.
\STATE  Update the virtual queues \(\{{X}_k(t+1),{Y}_k(t+1): k\in \mathcal{K}\}\) using the optimal \([\mathcal{P}^{(t)}]^*\) according to (\ref{backlog_x}) and (\ref{backlog_y}).
\ENDFOR
\STATE {\bf Output}: The policy for each time slot \([\mathcal{P}^{(t)}]^*\).
\label{Alg1:Iteration_End}
\end{algorithmic}
\end{algorithm}

\section{Simulation Results}
In this section, simulation results are provided to verify the accuracy of closed-form expressions and the performance of our proposed scheduling and power control approach. 
\subsection{Simulation Setup}
We consider a large square hall of \(50\times 50\) \({\rm m}^2\) with wrapped-around to avoid boundary effects. \(L=100\) APs are placed on the $h_{\text{AP}}=7$m high ceiling to form a uniform square array. \(K=200\) sensors with height $h_{\text{s}}=1.65$m are randomly distributed in this area. We model the large scale fading \(\beta_{l,k}\) as
\[
\beta_{l,k}=\mathcal{L}_{l,k}10^{\frac{{\sigma_{sh}z_{l,k}}}{10}}
\]
where \(10^{\frac{{\sigma_{\rm sh}z_{l,k}}}{10}}\) denotes the shadow fading with \(\sigma_{sh}=8\)dB and \(z_{l,k}\sim \mathcal {CN}(0,1)\), and the path loss \(\mathcal{L}_{l,k} (\rm dB)\) is given by
{\small\begin{align}\label{eq:ploss}
&\mathcal{L}_{l,k}
=& \left\{
\begin{array}{l}
  -\mathcal{L}_0 - 35\log_{10} (d_{l,k}), ~ \text{if} ~ d_{l,k}>d_1,\\
  -\mathcal{L}_0 - 15\log_{10} (d_1) - 20\log_{10} (d_{l,k}), ~ \text{if} ~ d_0< \!d_{l,k}\leq d_1,\\
  -\mathcal{L}_0 - 15\log_{10} (d_1) - 20\log_{10} (d_{0}), ~ \text{if} ~ d_{l,k} \leq d_0,\\
\end{array}%
\right.\nonumber
 \end{align}}
where \(d_0=10\)m, \(d_1=50\)m, and
\begin{align}
\mathcal{L}_0&\triangleq 46.3+33.9\log_{10}(f)-13.82\log_{10}(h_{\text{AP}})\nonumber\\
&-
(1.1\log_{10}(f)-0.7)h_{\text{s}}+(1.56\log_{10}(f)-0.8),\nonumber
\end{align}
with carrier frequency $f=1900$ MHz. The noise power is
\[
\sigma^2 = {B}\times k_B\times T_0\times \kappa,
\]
where \(k_B= 1.381\times 10^{-23}{\rm J/K}\), \(T_0=290K\), \(\kappa=9\)dB and the bandwidth \(B=20\) MHz. The other simulation parameters are summarized in Table I. In addition, The large scale fading \(\{\beta_{l,k},\forall l,k\}\) are generated once and fixed for all simulations.
\begin{table}[t!]
\renewcommand{\arraystretch}{1.3}
\caption{Simulation Parameters}
\label{table_parameters} \vskip-2mm
\centering
\begin{tabular}{|c|c|c|}
\hline
\bfseries parameter & \bfseries Meaning &\bfseries Value\\
\hline $L$ & Number of APs & 100\\
\hline $N$ & Number of antennas of each AP & 10\\
\hline $K$ & Number of all sensors & 200\\
\hline $B$ & Bandwidth & 20 MHz\\
\hline $\Delta/T_c$ & Coherence time& 0.2 s/200 symbols\\
\hline $\tau$ & Length of pilot & 60 symbols\\
\hline {$\rho_p$} & {Pilot transmit power} &  0.2 mW \\
\hline $\rho_u$ & Maximum uplink transmit power  & 20 mW \\
\hline $\zeta$ & Energy conversion efficiency  & 1 \\
\hline $\rho_d$ & Maximum downlink transmit power  & 20 W \\
\hline $\delta^{(t)}$ & Transmission model & Optimized \\
\hline ${\pmb \theta}^{(t)}$ & Active sensor set & Optimized \\
\hline ${\pmb \eta}^{(t)}$ & Downlink power control coefficients & Optimized \\
\hline ${\pmb \xi}^{(t)}$ & Uplink power control coefficients & Optimized \\
\hline $b_{\max}$ & Capacity of the battery & 300$mJ$ \\
\hline $b_0$ & Predefined threshold & 10 $mJ$ \\
\hline
\end{tabular}
\end{table}

\subsection{Accuracy of Expressions}
 \begin{figure}
	\centering \scalebox{1}{\includegraphics[width=\columnwidth]{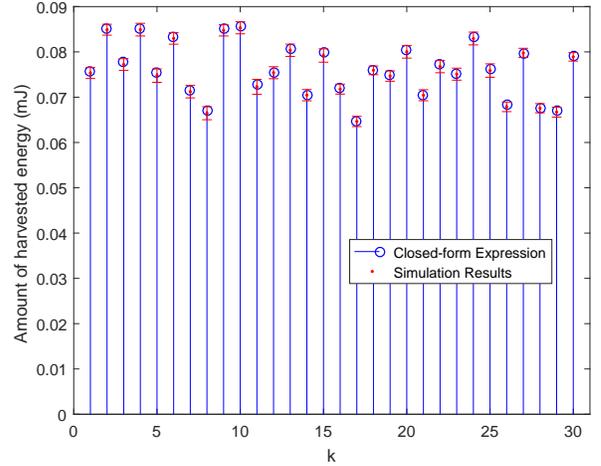}}
	\centering \caption{Accuracy of the amount of harvested energy \(\bar{\mathcal E}_{k}^{(t)}\) given in (\ref{Energy_Harvest_APP}).}
\end{figure}
Through a realization with $K_a=30$ sensors randomly scheduled, the accuracy of the closed-form expressions \(\bar{\mathcal E}^{(t)}_{k}\) in (\ref{Energy_Harvest_APP}) and \(\bar{R}^{(t)}_{k}\) in (\ref{Achievablerates_APP}) are verified in Fig. 2 and Fig. 3, respectively. In Fig 2, the closed-form expressions \(\bar{\mathcal E}_{k}\) independent of random pilots are compared with the simulation results obtained through 500 realizations of random pilot sequences with the uniform power control, i.e., \({\eta}^{(t)}_{l,k}\gamma^{(t)}_{l,k}=1/K_a, \forall l,~\mbox{and}~ k\in {\mathcal{K}_a^{(t)}}\). The figure shows that the closed-form expressions agree well with the mean of simulation results. In addition, the small variances of simulation results stemming from random pilots reveals that pilot optimization over random pilots is not necessary.

In Fig. 3, the closed-form expressions \(\bar{R}^{(t)}_{k}\) independent of random pilots are compared with the simulation results obtained through 500 realizations of random pilot sequences with uniform power control, i.e., \(\xi^{(t)}_k=1, k\in \mathcal{K}^{(t)}_a\). Similarly as in Fig. 2, it can be seen that the difference between the closed-form expressions and the simulation results is small. 

To investigate the impact of enlarging the active set $\mathcal{K}_a^{(t)}$, the average \(\bar{\mathcal E}_{k}^{(t)}\) and \(\bar{R}_{k}^{(t)}\) obtained through 500 random schedule realizations versus $K_a$ is plotted in Fig. 4. As noted in Remark 1, the metrics \(\bar{\mathcal E}_{k}^{(t)}\) and \(\bar{R}_{k}^{(t)}\) decrease as $K_a$ increases. In addition, it can be seen that \(\bar{\mathcal E}_{k}^{(t)}\) is more sensitive to $K_a$, which implies the importance of scheduling during WPT.

 \begin{figure}
	\centering \scalebox{1}{\includegraphics[width=\columnwidth]{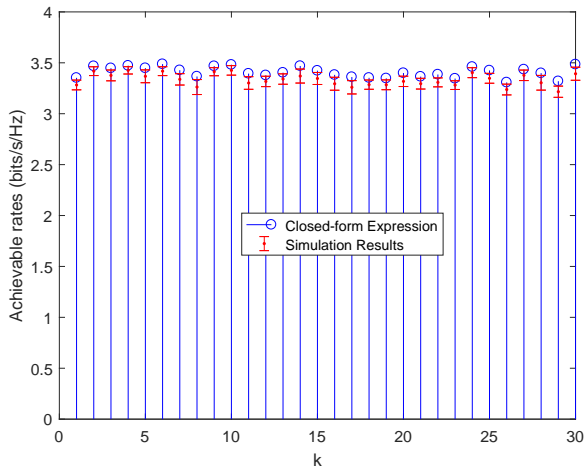}}
	\centering \caption{Accuracy of the achievable rate \(\bar{R}_{k}^{(t)}\) given in (\ref{Achievablerates_APP}).}
\end{figure}

 \begin{figure}
	\centering \scalebox{1}{\includegraphics[width=\columnwidth]{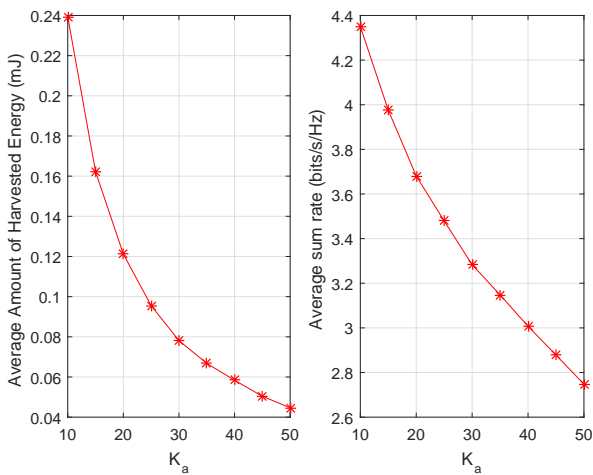}}
	\centering \caption{Average amount of harvested energy and average achievable rate versus the number of scheduled sensors \(K_a^{(t)}\).}
\end{figure}

\subsection{Performance Comparison}
In this subsection, we evaluate the performance of our proposed long-term scheduling and power control approach. For comparison, we consider the following simple greedy scheme as a benchmark. In each time slot for transmission mode, $K_a$ sensors with the largest $b_k^{(t)}$ are scheduled for data transmission with \(\xi^{(t)}_k=1, k\in \mathcal{K}^{(t)}_a\). The transmission mode continues until there exits some sensors whose battery is depleted, i.e., $b_k^{(t)}=0$, and then the harvesting mode is triggered. In each time slot for harvesting, the $K_a$ sensors with the lowest $b_k^{(t)}$ are scheduled for WPT with the uniform power allocation, i.e., \({\eta}_{l,k}^{(t)}\gamma_{l,k}^{(t)}=1/K_a, \forall l,~\mbox{and}~ k\in {\mathcal{K}_a^{(t)}}\). Until $b_k^{(t)}\geq b_0, k\in \mathcal{K}$, the transmission mode is triggered again. In our simulations, we consider $K_a=30$ and $100$. 

Fig. 5 and Fig. 6 plot the dynamic of the minimum time average rate with $K_a=30$ and 100, respectively. It can be seen that the minimum time average rate becomes stable after about 1000 time slots. The shadowed error bar is $$\hat{\sigma}^{(T)}=\sqrt{\frac{\sum_{k=1}^K[\frac{1}{T}\sum_{t=0}^{T-1}R_k^{(t)}-\overline{R^{(T)}}]^2}{K}},$$ where $\overline{R^{(T)}}=\{\frac{1}{T}\sum_{t=0}^{T-1}\sum_{k=1}^KR_k^{(t)}\}/K$ is the mean of time average rates over all sensors. The small shadowed error bar reflects the max-min fairness of our proposed approach. Compared with the greedy benchmark, our proposed approach can boost the minimum time average rate significantly. The improvement mainly results from two aspects: On one hand, the power consumption for data transmission is significantly reduced since (\ref{P3B.1}) strikes a balance between the spectrum efficiency and the power consumption, instead of focusing only on the spectrum efficiency. On the other hand, the WPT efficiency is improved through optimizing the downlink power control coefficients and scheduling. Moreover, it is seen that a larger $W$ leads to a higher minimum time average rate, and requires more time slots to achieve the rate stability. Comparing Fig. 5 and Fig. 6, the minimum time average rate becomes smaller as $K_a$ increased from 30 to 100 although more sensors are active in each time slot. This is because the accuracy of channel estimation is reduced by enlarging the active set $\mathcal{K}_a^{(t)}$, which  is noted in Remark 1. 
 \begin{figure}
	\centering \scalebox{1}{\includegraphics[width=\columnwidth]{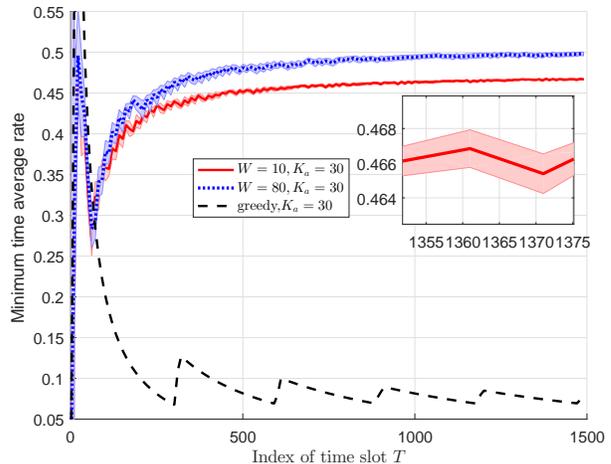}}
	\centering \caption{The minimum time average rate versus the index of time slot with $K_a=30$.}
\end{figure}
\begin{figure}
	\centering \scalebox{1}{\includegraphics[width=\columnwidth]{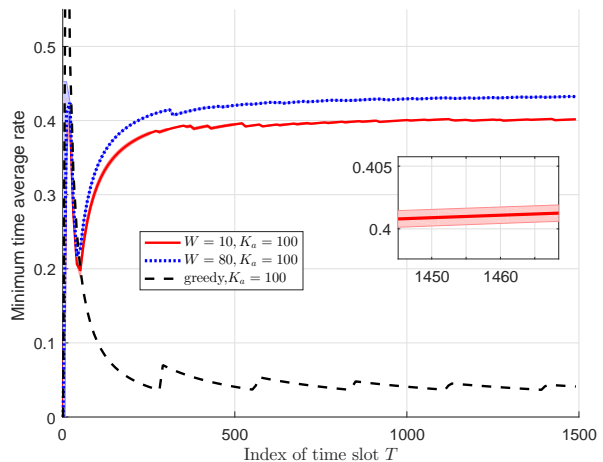}}
	\centering \caption{The minimum time average rate versus the index of time slot with $K_a=100$.}
\end{figure}

To investigate the rate stability of the virtual queues, Fig. 7 shows the sum of time average backlogs $\overline{X}(t)$ and $\overline{Y}(t)$ versus the index of time slot with different \(W\), where  $$\overline{X}(t)=\frac{1}{T}\sum_{t=0}^{T-1}\sum_{k=1}^KX_k(t),~\mbox{and}~ \overline{Y}(t)=\frac{1}{T}\sum_{t=0}^{T-1}\sum_{k=1}^KY_k(t).$$ The non-increasing $\overline{X}(t)$ and $\overline{Y}(t)$ reveal that $\{X_k(t),Y_k(t): k\in \mathcal{K}\}$ are rate stable, which implies the long-term constraints (\ref{P2.1}) and (\ref{long_term_cons}) of problem ${\bf P}_2$ are satisfied. With the increase of \(W\), the time average backlogs $\overline{X}(t)$ and $\overline{Y}(t)$ also increase. Fig. 8 plots the dynamic of the battery state $\{b_k^{(t)}:k\in \mathcal{K}\}$ versus the index of time slot. It can be seen $b_k^{(t)}$ always fluctuates across the predefined $b_0$, which implies that the batteries are never exhausted. If $b_k^{(t)}$ is smaller than $b_0$ for some continuous time slots, the corresponding backlog $X_k{(t)}$ would increase continuously, and then trigger power transfer to increase $b_k^{(t)}$ and avoid the depletion of the battery.
 \begin{figure}
	\centering \scalebox{1}{\includegraphics[width=\columnwidth]{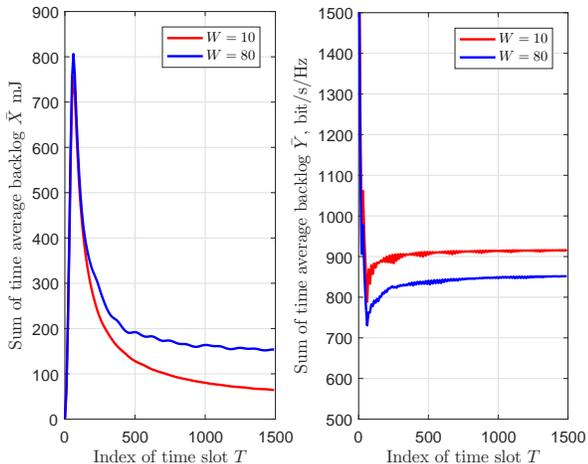}}
	\centering \caption{The backlogs versus the index of time slot with different \(W\).}
\end{figure}
 \begin{figure}
	\centering \scalebox{1}{\includegraphics[width=\columnwidth]{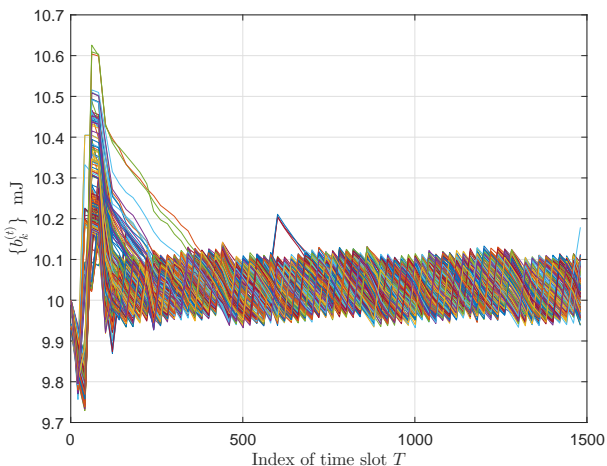}}
	\centering \caption{The dynamic of $\{b_k^{(t)}: k\in \mathcal{K}\}$ with \(W=10\).}
\end{figure}

\section{Conclusions}
In this paper, we have considered the long-term scheduling and power control in a wirelessly powered cell-free IoT network. We first derived closed-form expressions for the harvested energy and the achievable rates, and then formulated a long-term scheduling and power control problem to maximize the minimum time average achievable rate. Following the Lyapunov optimization approach, the transmission mode, the sensor state, the uplink and downlink power control coefficients are jointly determined for each time slot. Simulation results reveal that the proposed long-term scheduling and power control approach can boost the max-min time average achievable rate significantly.

\vspace{0.2in}
\section*{Appendices}

\subsection{Proof of Theorem 1}
To prove Theorem 1, we first introduce the following two lemmas.
\begin{lemma}[\cite{J. Hoydis}] \label{lemma5}
Let ${\pmb A}\in \mathbb{C}^{\tau\times \tau}$ is a Hermitian invertible matrix. Then, for any vector ${\pmb x}\in \mathbb{C}^{\tau}$ and any scalar $a\in \mathbb{C} $ such that ${\pmb A}+a{\pmb x}{\pmb x}^H$ is invertible,
\begin{align}\label{matrix_inverse}
   {\pmb x}^H\left({\pmb A}+a{\pmb x}{\pmb x}^H\right)^{-1}=\frac{{\pmb x}^H{\pmb A}^{-1}}{1+a{\pmb x}^H{\pmb A}^{-1}{\pmb x}}. 
\end{align}
\end{lemma}

\begin{lemma}[\cite{J. Hoydis}] \label{lemma6}
Let ${\pmb A}\in \mathbb{C}^{\tau\times \tau}$, and  ${\pmb x},{\pmb y}\sim \mathcal{CN}({\pmb 0}, \frac{1}{\tau}{\pmb I}_{\tau})$. Assume that ${\pmb A}$ has uniformly bounded spectral norm (with respect to $\tau$) and that ${\pmb x}$ and ${\pmb y}$ are mutually independent and independent of ${\pmb A}$, we have
\begin{align}
   & {\pmb x}^H{\pmb A}{\pmb x}-{{\rm tr}{\pmb A}}/{\tau}\mathop{-\!\!\!\!-\!\!\!\!-\!\!\!\!-\!\!\!\!-\!\!\!\!-\!\!\!\!\longrightarrow}\limits_{\tau\rightarrow\infty}^{{\rm a}.{\rm s}.}0, \label{xAx}\\
    &{\pmb x}^H{\pmb A}{\pmb y}\mathop{-\!\!\!\!-\!\!\!\!-\!\!\!\!-\!\!\!\!-\!\!\!\!-\!\!\!\!\longrightarrow}\limits_{\tau\rightarrow\infty}^{{\rm a}.{\rm s}.}0\label{xAy},
\end{align}
\end{lemma}
Substituting \(\pmb{Z}^{(t)}_{l,k}=\pmb{Z}^{(t)}_l-E_p\beta_{l,k}\pmb{\psi}_k\pmb{\psi}_k^H\) into (\ref{Gamma_lk}), and using (\ref{matrix_inverse}), we obtain
	\begin{flalign*}
		\gamma_{l,k}^{(t)}
		&={E_p}\beta^2_{l,k}\pmb{\psi}_k^H\left(\pmb{Z}^{(t)}_{l,k}+E_p\beta_{l,k}\pmb{\psi}_k\pmb{\psi}_k^H\right)^{-1}\pmb{\psi}_k\\
		&=\frac{{E_p}\beta^2_{l,k}\pmb{\psi}_k^H(\pmb{Z}_{l,k}^{(t)})^{-1}\pmb{\psi}_k}{1+{E_p}\beta_{l,k}\pmb{\psi}_k^H(\pmb{Z}_{l,k}^{(t)})^{-1}\pmb{\psi}_k}.
		\end{flalign*}
Since \(\pmb{\psi}_k \sim \mathcal{CN}\left( 0, \frac{1}{\tau}{\pmb I}\right)\) is independent of $[\pmb{Z}_{l,k}^{(t)}]^{-1}$, using (\ref{xAx}), we have	
	\begin{flalign}\label{APP.A.1}
	{\gamma}_{l,k}^{(t)}-\frac{{E_p}\beta^2_{l,k}{\rm tr}[(\pmb{Z}_{l,k}^{(t)})^{-1}]/\tau}{1+{E_p}\beta_{l,k}{\rm tr}[(\pmb{Z}_{l,k}^{(t)})^{-1}]/\tau}
	\xrightarrow[]{\text{a.s.}}0,
		\end{flalign}
Substituting (\ref{Lemma1.A}) into (\ref{APP.A.1}), we conclude the proof. \hfill{\(\blacksquare\)}

\subsection{Proof of Theorem 2}
Since \(\pmb{a}^{(t)}_{l,k}=\sqrt{E_p}\beta_{l,k}[\pmb{Z}^{(t)}_l]^{-1}\pmb{\psi}_k\), using (\ref{matrix_inverse}), (\ref{xAx}) and (\ref{xAy}), we obtain 
\begin{align}
    &\left\|\pmb{a}^{(t)}_{l,k}\right\|^2-\frac{E_p\beta^2_{l,k}{\rm tr}\{[\pmb{Z}_{l,k}^{(t)}]^{-2}\}/\tau}{\left(1+E_p\beta_{l,k}{\rm tr}\{[\pmb{Z}_{l,k}^{(t)}]^{-1}\}/\tau\right)^2}\xrightarrow[]{\text{a.s.}}0, \label{APP.C.1}\\
    &\pmb{\psi}_k^H\pmb{a}^{(t)}_{l,k}-\frac{\sqrt{E_p}\beta_{l,k}{\rm tr}\{[\pmb{Z}_{l,k}^{(t)}]^{-1}\}/\tau}{1+E_p\beta_{l,k}{\rm tr}\{[\pmb{Z}_{l,k}^{(t)}]^{-1}\}/\tau}\xrightarrow[]{\text{a.s.}}0,\label{APP.C.2}
\end{align}
and 
\begin{align}\label{APP.C.3}
    \pmb{\psi}_i^H\pmb{a}_{l,k}^{(t)}\xrightarrow[]{\text{a.s.}}0, \quad (i\neq k).
\end{align}
 Substituting (\ref{Lemma1.A}), (\ref{Lemma1.B}), (\ref{gamma_app}), (\ref{APP.C.1}), (\ref{APP.C.2}) and (\ref{APP.C.3}) into (\ref{Per_B.3}), we can conclude the proof. \hfill{\(\blacksquare\)}
 
\subsection{Proof of Lemma 3}
According to (\ref{mean_error}) and (\ref{APP.A.1}), we have  
\begin{flalign}
		&\bar{e}_{l,k}^{(t)}-\frac{\beta_{l,k}}{1+{E_p}\beta_{l,k}{\rm tr}[(\pmb{Z}_{l,k}^{(t)})^{-1}]/\tau}
	\xrightarrow[]{\text{a.s.}}0,\label{bar_gamma_lk_1}\\
		&[\bar{e}_{l,k}^{(t)}]'-\frac{\beta_{l,k}}{1+{E_p}\beta_{l,k}{\rm tr}\{[({\pmb{Z}}_{l,k}^{(t)})']^{-1}\}/\tau}
	\xrightarrow[]{\text{a.s.}}0. \label{bar_gamma_lk_2}
		\end{flalign}
	 Without loss of generality, we assume \[[{\pmb{Z}}^{(t)}_{l,k}]'=\pmb{Z}_{l,k}^{(t)}+E_p\beta_{j}\pmb{\psi}_{l,j}\pmb{\psi}_{j}^H,j\in [\mathcal{K}_a^{(t)}]', j\notin \mathcal{K}_a^{(t)}.\] Using (\ref{matrix_inverse}), we have
	\begin{align}
	    {\rm tr}\{[({\pmb{Z}}_{l,k}^{(t)})']^{-1}\}
	     &={\rm tr}\{[\pmb{Z}_{l,k}^{(t)}]^{-1}\}-\frac{E_p\beta_{l,j}\pmb{\psi}_j^H[\pmb{Z}_{l,k}^{(t)}]^{-2}\pmb{\psi}_j}{1+E_p\beta_{l,j}\pmb{\psi}_j^H[\pmb{Z}_{l,k}^{(t)}]^{-1}\pmb{\psi}_j}\nonumber\\
	      &\xrightarrow[]{\text{a.s.}}{\rm tr}\{[\pmb{Z}_{l,k}^{(t)}]^{-1}\}-\frac{E_p\beta_{l,j}{\rm tr}\{[\pmb{Z}_{l,k}^{(t)}]^{-2}\}}{1+E_p\beta_{l,j}{\rm tr}\{[\pmb{Z}_{l,k}^{(t)}]^{-1}\}}\nonumber\\
	&<{\rm tr}\{[\pmb{Z}_{l,k}^{(t)}]^{-1}\}.\label{TR_Z_Lk}
	\end{align}
	where the second step follows from (\ref{xAx}) due to the independence of $\pmb{\psi}_{j}$ and $[\pmb{Z}_{l,k}^{(t)}]^{-1}$. Substituting (\ref{TR_Z_Lk}) into (\ref{bar_gamma_lk_1}) and (\ref{bar_gamma_lk_2}), we can conclude the proof.\hfill{\(\blacksquare\)}

\subsection{Proof of Theorem 3}
 According to the definition \([\mathcal{P}^{(t)}]^{*}\) in (\ref{Sec4.11a}), we have
\begin{align}\label{APP_D.1}
\mathcal{D}^{*}(t)\leq\mathcal{D}'(t)
\end{align}
 at each time slot \(t=0,1,\cdots\). From the identical initial point \(X_k^*(0)=X_k'(0)=X_k(0), Y_k^*(0)=Y_k'(0)=Y_k(0), k=1,\cdots,K\),  using (\ref{drift_def}) and (\ref{APP_D.1}), we can obtain 
 \begin{align}\label{APP_D.2}
&\mathcal{L}'(1)=\mathcal{D}'(0)+\sum_{k=1}^K\{[X_k'(0)]^2+[Y_k'(0)]^2\}\nonumber\\
&\geq\mathcal{D}^*(0)+\sum_{k=1}^K\{[X_k(0)]^2+[Y_k(0)]^2\}=\mathcal{L}^*(1),
\end{align}
where $\mathcal{L}'(t)=\sum_{k=1}^K\{[X_k'(t)]^2+[Y_k'(t)]^2\}$ and $\mathcal{L}^*(t)=\sum_{k=1}^K\{[X_k^*(t)]^2+[Y_k^*(t)]^2\}$. For any given time slot \(t\) with $\mathcal{L}^*(t) \leq \mathcal{L}'(t)$, using (\ref{drift_def}) and (\ref{APP_D.1}), we have
 \begin{align}\label{APP_D.3}
     &\mathcal{L}^*(t+1)-\mathcal{L}^*(t)\leq \mathcal{L}'(t+1)-\mathcal{L}'(t) \nonumber\\
     &= \sum_{k=1}^K\{[X_k'(t+1)]^2+[Y_k'(t+1)]^2\}-\sum_{k=1}^K\{[X_k'(t)]^2+[Y_k'(t)]^2\} \nonumber\\
     &\leq \mathcal{L}'(t+1)-\mathcal{L}^*(t) \nonumber
 \end{align}
 which implies 
 \begin{align}
     \mathcal{L}^*(t+1) \leq \mathcal{L}'(t+1).
 \end{align}
From (\ref{APP_D.2}) and (\ref{APP_D.3}), we can conclude that 
\begin{align}\label{APP_D.4}
\sum_{k=1}^K\{[X_k^*(T)]^2+[Y_k^*(T)]^2\}\leq \sum_{k=1}^K\{[X_k'(T)]^2+[Y_k'(T)]^2\} 
\end{align} for any given \(T\). When $\{X'_k(t),Y'_k(t):k=1,\cdots, K,\}$ are rate stable, (\ref{APP_D.4}) is equivalent to 
 \begin{align}\label{APP_D.5}
\mathop{\lim}_{T\rightarrow \infty}\frac{1}{T}\sum_{k=1}^K\{[X_k^*(T)]^2+[Y_k^*(T)]^2\}\leq 0,
\end{align}
 which implies $\{X^*_k(t),Y^*_k(t):k=1,\cdots, K,\}$ are rate stable. Then, we conclude the proof. \hfill{\(\blacksquare\)}

\end{document}